\documentclass[epsfig,psfig,aps,twocolumn,prb,showpacs]{revtex4-2}
\usepackage[colorlinks=true,citecolor=red,urlcolor=blue]{hyperref}
\usepackage{amsfonts}
\usepackage{graphicx}
\usepackage{epsfig}
\usepackage{epsf}
\usepackage{amsmath}
\usepackage{physics}
\usepackage{epstopdf}
\usepackage{float} 
\usepackage{amstext}
\usepackage{bm}
\usepackage{mathrsfs}
\newcommand{\be}{\begin{equation}}
\newcommand{\beqn}{\begin{eqnarray}}
\newcommand{\eeq}{\end{equation}}
\newcommand{\ee}{\end{equation}}
\begin{document}

\title{Klein tunneling in deformed honeycomb-dice lattice: from massless to massive particles} 
\author{L. Mandhour}
\email{lassaad.mandhour@istmt.utm.tn}
\author{F. Bouhadida}

\affiliation{Laboratoire de Physique de la Mati\`ere Condens\'ee, Facult\'e des Sciences de Tunis, Universit\'e de Tunis el Manar, Campus Universitaire Tunis, El Manar, 2092 Tunis, Tunisie.}

\begin{abstract}
 
We show that under compressive uniaxial deformation of the three-band $\alpha-T_3$ lattice, the Dirac cones move toward each other, merge, and a gap opens, while the flat band remains unchanged. Consequently, the low-energy spectrum transitions from linear to quadratic dispersion, indicating the shift from massless to massive Dirac particles.
Here, we theoretically investigate the tunneling properties of particles through a sharp $\mathit{np}$ junction in a deformed $\alpha-T_3$ lattice, focusing on the case where the particle energy is half the junction height.
 We show that this transition from massless to massive particles leads to a change from omnidirectional total transmission, known as super-Klein tunneling, to omnidirectional total reflection, referred to as anti-super-Klein tunneling, in the case of the dice lattice ($\alpha=1$). For all values of $\alpha$, this transition manifests as a change from conventional Klein tunneling to anti-Klein tunneling.

\end{abstract}

\date{\today}
\maketitle

\section{Introduction}

Klein tunneling (KT) is characterized by the perfect transmission of relativistic particles incident perpendicular to a potential barrier whose height exceeds twice the particle's rest energy, $mc^2$ (where $m$ is the mass and $c$ is the speed of light) \cite{Klein1929, Calogeracos, Dombey}.  
In particle physics, the experimental realization of KT is practically unattainable due to the enormous electric fields required.
However, the discovery of graphene \cite{Novoselov}, a single layer of carbon atoms arranged in a honeycomb lattice (HCL), has enabled the experimental realization of KT in condensed matter systems \cite{Huard, Stander, Young2009}.
This is because charge carriers in graphene behave as chiral, massless Dirac fermions with pseudospin-1/2, which is conserved across the barrier interface \cite{Katsnelson, Allain}.

KT is not limited to graphene; it also occurs in relativistic materials described by the Dirac-Weyl equation with enlarged pseudospin $S>1/2$ \cite{Zhihao}.
In particular, pseudospin-1 systems exhibit a remarkable transport phenomenon known as super-Klein tunneling (SKT), characterized by omnidirectional perfect transmission when the particle energy equals half of the junction height \cite{Urban, Betancur, Fang, Shen, Wu2024}.
SKT has also been reported in systems of spinless Klein–Gordon particles \cite{Kim2019} and in pseudospin-1/2 Dirac materials \cite{Betancur2018}.

The dice or $T_3$ lattice \cite{Sutherland,Vidal,Vidal2001}, an example of pseudospin-1 system, presents the same structure as HCL but with an additional site $C$ at the center of each hexagon, related to one of the $A$ or $B$ sites. The low-energy behavior of the dice lattice is governed by the Dirac-Weyl Hamiltonian, as in graphene, but with pseudospin $S = 1$.  Its corresponding low-energy band structure resembles that of the HCL, with an additional flat band at the Dirac points.
The $\alpha-T_3$ model, introduced by A. Raoux \textit{et al.} \cite{Raoux}, interpolates between the HCL ($\alpha=0$) and the dice lattice ($\alpha=1$). In this lattice, the low-energy excitations are described by a Dirac–Weyl Hamiltonian with a hybrid pseudospin $S = 1/2-1$ \cite{Illes2016}. 
It has been shown that $Hg_{1-x}Cd_xTe$ with a critical value of $x=0.17$, maps onto the $\alpha-T_3$ model for $\alpha = \frac{1}{\sqrt{3}}$ \cite{Malcolm}.
E. Illes \textit{et al.} \cite{Illes2017} studied KT in the $\alpha-T_3$ model and have found a perfect transmission at normal incidence for all values of $\alpha$, with enhanced transmission at other angles as $\alpha$ increases.
Many other tunneling properties have also been explored in the $\alpha-T_3$ model, including Dirac particle tunneling through a periodic potential barrier \cite{Cunha2022}, a potential barrier under a linearly polarized off-resonant dressing field \cite{Iurov2022}, and combined electric and magnetic barriers \cite{Bouhadida2020}.

In contrast to monolayer graphene, where particles behave as chiral massless Dirac fermions, the charge carriers in bilayer graphene are chiral massive Dirac fermions and exhibit total reflection at normal incidence on a potential barrier. This phenomenon, known as anti-Klein tunneling (AKT), arises from pseudospin conservation between the incident and reflected particles \cite{Katsnelson, Allain, Varlet2016, Du}.
AKT has also been reported in other relativistic materials such as deformed honeycomb lattices \cite{Bahat}, graphene with strong Rashba spin-orbit coupling \cite{Liu2016} and semi-Dirac materials \cite{Banerjee}. 
The counterpart of SKT in pseudospin-1 systems is omnidirectional total reflection, referred to as anti-super-Klein tunneling (ASKT). This effect, observed in phosphorene, arises from the fact that the pseudospins of the incident and transmitted electrons are antiparallel \cite{Yonatan}.

Under a uniaxial deformation of the honeycomb lattice (HCL), the Dirac points are shifted away from the corners of the first Brillouin zone.
For a sufficiently strong deformation, the two Dirac points merge into a single one, leading to the opening of a gap. This merging of Dirac points signals a topological transition from a semi-metallic to an insulating phase \cite{Gilles1,Pereira,Gail2009,Gilles2,Tarruel,Feilhauer}. 
At the merging point, known as the semi-Dirac point, the dispersion is linear along one direction and quadratic along the other \cite{Gilles1,Gilles2}. 
In a tight-binding picture, the motion and eventual merging of Dirac points can be achieved by tuning one of the three nearest-neighbor hopping parameters in the HCL \cite{Gilles1}. This mechanism, however, is not accessible in graphene due to its high stiffness \cite{Pereira}. 
Nevertheless, the merging of Dirac points can be realized in artificial graphene systems \cite{Polini2013}, such as ultracold atoms trapped in a honeycomb optical lattice \cite{Tarruel}, microwave photonic crystals \cite{Bellec} and for a review see \cite{Montambaux2018}. More recently, it has been experimentally explored inside bulk ZrSiS under in-plane magnetic fields \cite{Shao2024}.
 Another type of deformation in graphene, known as the Kekulé distortion, can also induce the merging of the Dirac points at the center of the Brillouin zone, and has been experimentally observed in graphene grown on a copper substrate \cite{Gutierrez2016}. 
Depending on the distortion pattern, it either opens a gap (Kek-O phase) or produces a pair of superimposed Dirac cones with different Fermi velocities (Kek-Y phase) \cite{Iurov2023, Mojarro2020}.

The impact of honeycomb lattice (HCL) deformation on electronic transport has been extensively investigated \cite{Bahat, Banerjee, Betancurprb2018, Saha, Adroguer, Shengyuan}.
A transition from KT to AKT has been reported in deformed HCL, either by tuning the strength \cite{Bahat} or the direction \cite{Banerjee} of the deformation. This transition is associated with the nature of particles which evolve from massless to massive Dirac fermions.
The KT to AKT transition has also been studied in other systems, such as double-Weyl semimetals \cite{Zhu2020} and bilayer graphene \cite{Anna,Du,Donck}.  
To the best of our knowledge, however, the transition from SKT to ASKT has not yet been explored. 

In this paper, we theoretically study KT across an $np$ junction in a deformed $\alpha-T_3$ lattice, focusing on the effects of the deformation, the parameter $\alpha$, and the orientation of the $np$ junction relative to the deformation axis.
We identify three distinct phases under a continuous compressive uniaxial deformation of the $\alpha-T_3$ lattice.
In the first, the Dirac phase, the tunneling properties across the $\mathit{np}$ junction, such as KT, SKT and the $\alpha$-dependent junction transparency, are similar to those reported in the undeformed $\alpha-T_3$ model. In the second, the intermediate phase, intervalley scattering destroys both KT and SKT. In the third phase, when the Dirac cones merge, ASKT is observed when the junction is parallel to the deformation direction, whereas SKT occurs when the junction is perpendicular to the deformation direction.  
A transition from the SKT to the ASKT in the dice lattice ($\alpha=1$) can be realized either by rotating the junction at the merging point or by applying a continuous uniaxial deformation parallel to the junction.
Similarly, the transition from the KT to AKT is observed for all values of the parameter $\alpha$.

The paper is organized as follows. In Section \ref{II}, we introduce the deformed $\alpha-T_3$ model and discuss the effects of uniaxial deformation on the energy spectrum, wave functions, and topological properties.
Section \ref{III} presents the effect of deformation on tunneling across an $np$ junction oriented parallel to the deformation direction. 
In Section \ref{IV}, we examine the impact of the $\mathit{np}$ junction orientation relative to the deformation axis in the semi-Dirac phase.
Finally, conclusions are drawn in Section \ref{V}.

\section{The deformed $\alpha-T_3$ lattice} \label{II}
\subsection{Low-energy band dispersion}
Graphene is characterized by a honeycomb lattice (HCL) consisting of two inequivalent sites, labeled $A$ and $B$, connected by a hopping amplitude $t$. Starting from this HCL, the dice lattice ($T_3$ lattice) is constructed by adding a third site, $C$, at the center of each hexagon. Each $C$ site is connected to one of the two inequivalent sites (for example, $B$) with a hopping amplitude $t_{BC} = t_ {AB} = \frac{t}{\sqrt{2}}$.

In the $\alpha-T_3$ lattice,  the site $B$ is coupled to the three $A$ sites via the hopping amplitude $t_{AB}^{\vec{\delta_i}}=t \cos\varphi$ and to the three $C$ sites via $t_{BC}^{-\vec{\delta_i}}=t \sin\varphi$, where $\vec{\delta_i}(i=1,2,3)$ are the vectors connecting the nearest-neighbor sites. For convenience, we introduce the parameter $\varphi$ given by $\tan \varphi=\alpha$.

\begin{figure}[h!]
  \centering
\setlength{\unitlength}{1mm}
\includegraphics[width=0.45\textwidth]{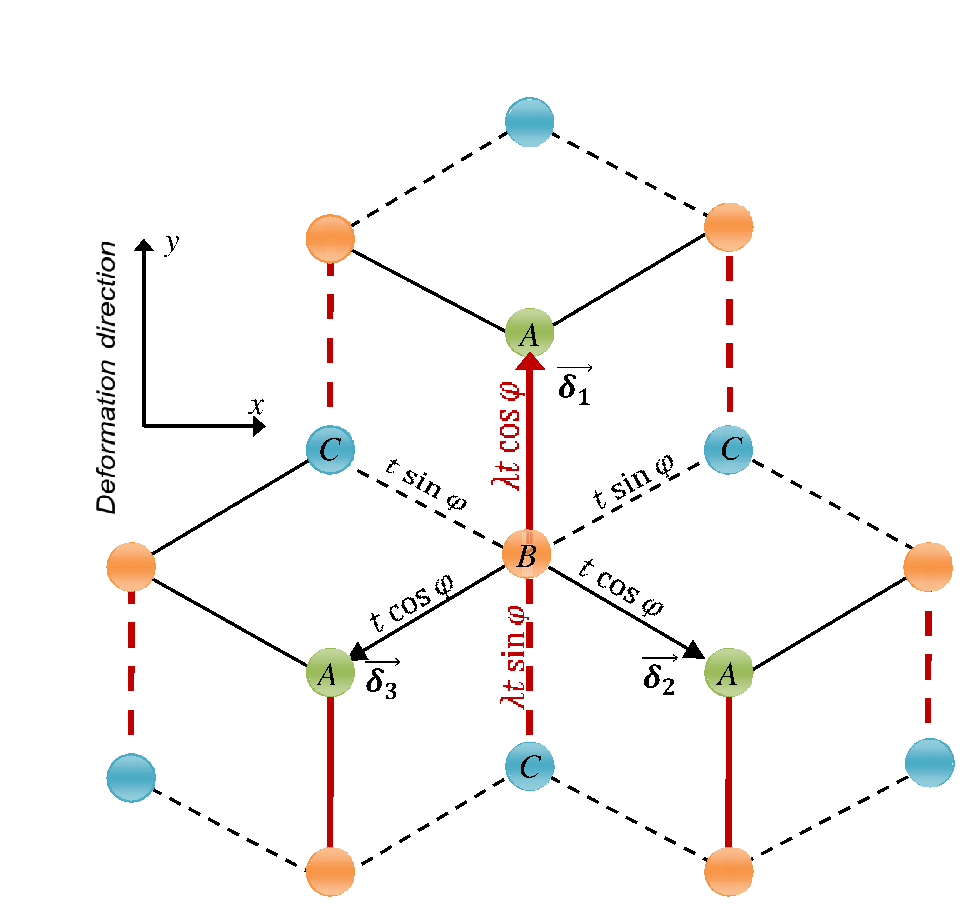}

\caption{(Color online) Schematic representation of the deformed $\alpha-T_3$ lattice. Each unit cell contains three sites, $A$, $B$, and $C$.  The hopping amplitudes between $A$ and $B$ sites are $\lambda t \cos\varphi$ ($\lambda >1$) along the deformation direction (red thick lines) and $t \cos\varphi$ along the other directions (black thin lines). The hopping amplitude between $B$ and $C$ sites is $\lambda t \sin\varphi$ along the deformation direction (red dashed thick lines) and $ t \sin\varphi$ (black dashed thin lines) along the other directions.}
\label{figure1}
\end{figure}

When a compressive uniaxial deformation is applied along the $\vec{\delta_1}$ direction (the $y$ direction), the corresponding hopping amplitudes are modified as $t_{AB}^{'\vec{\delta_1}}= \lambda t_{AB}^{\vec{\delta_1}}$, $t_{BC}^{'-\vec{\delta_1}}= \lambda t_{BC}^{-\vec{\delta_1}}$ (see Fig.\ref{figure1}), where the parameter $\lambda\geq1$ quantifies the strength of the lattice deformation.

Following the nearest-neighbor tight-binding model, the electronic properties of the deformed $\alpha-T_3$ lattice are described by the Hamiltonian
\be
 H=\begin{pmatrix}
0 & \cos\varphi f_{\lambda}^{*}(\vec{k})  & 0\\ 
\cos\varphi f_{\lambda}(\vec{k}) & 0 &\sin\varphi f_{\lambda}^{*}(\vec{k}) \\ 
0& \sin\varphi f_{\lambda}(\vec{k}) & 0
\end{pmatrix}, 
\label{eq1} 
\ee
where 
\be
f_{\lambda}(\vec{k})=t\left(\lambda e^{i\vec{k} \vec{\delta_1}}+e^{i\vec{k} \vec{\delta_2}}+ e^{i\vec{k} \vec{\delta_3}}\right),
\label{fk}
\ee
with the nearest-neighbors vectors defined as $\vec{\delta_1}=a\vec{e_y} $, $\vec{\delta_2}=\frac{a}{2}\left (\sqrt{3}\vec{e_x}-\vec{e_y} \right )$,  $\vec{\delta_3}=\frac{a}{2}\left (-\sqrt{3}\vec{e_x}-\vec{e_y} \right )$ and $a$ denotes the intersite distance.

The energy spectrum of the deformed $\alpha-T_3$ lattice consists of a flat band at $E=0$ and two dispersive bands $E=s \mid f_{\lambda}(\vec{k})\mid$, where $s=\pm1$ is the band index.
The corresponding eigenfunctions are
\begin{subequations}
\be 
\Psi^{0} _{\lambda}(\vec{r})=\begin{pmatrix}
\sin\varphi e^{-i\theta_{\lambda} } \\ 
0\\ 
-\cos\varphi e^{i\theta_{\lambda} }
\end{pmatrix}e^{i\vec{k}\vec{r}},
\ee
\be 
\Psi^{s} _{\lambda}(\vec{r})= \frac{1}{\sqrt{2}} \begin{pmatrix}
\cos\varphi e^{-i\theta_{\lambda} } \\ 
s \\ 
\sin\varphi e^{i\theta_{\lambda} } 
\end{pmatrix}e^{i\vec{k}\vec{r}},
\ee 
\label{eq2}
\end{subequations}
where the phase is defined as $\theta_{\lambda} =\arg (f_{\lambda}(\vec{k}))$.\\
A notable feature of this model is that, for a given deformation strength $\lambda$, the energy spectrum is independent of $\alpha$, whereas the eigenfunctions do depend on it.

Let us now discuss the effect of the deformation on the $\alpha-T_3$ band structure.
For $\lambda =1 $, we recover the undeformed $\alpha-T_3$ lattice. 
In this case, the low-energy spectrum consists of linear bands touching at the $K$ and $K'$ points of the Brillouin zone, similar to graphene, but with an additional flat band at $E=0$.
When $1<\lambda<2$, the Dirac cones shift from the corners $K$ and $K'$ and move closer to each other along the $k_x$ direction as $\lambda$ increases.
The new positions of Dirac points are given by $D_\pm \left(\frac{2\pi}{\sqrt{3}a}\pm \frac{2}{\sqrt{3}a}\arccos\left(\lambda/2\right),0\right)$, determined from the condition $f_{\lambda}(\vec{k}_{D_\pm})= 0$. 
At $\lambda =2$, the Dirac points $D_\pm$ merge at the $M$ point, $M\left(\frac{2\pi}{a\sqrt{3}},0\right)$, and for $\lambda>2$, a gap opens in the spectrum. This evolution reflects a transition from massless Dirac fermions in the undeformed lattice to massive Dirac fermions in the gapped phase, directly influencing electronic properties such as tunneling through $\mathit{np}$ junctions, as we discuss in the next section.

Here, we focus on the low-energy Hamiltonian. In the vicinity of the Dirac points $D_\xi$ ($\xi=\pm$) for $1\leq \lambda\leq 2$, the function $f_{\lambda}(\vec{k})$ can be approximated as
\be
f_{\lambda}(\vec{k}_{D_\xi}+\vec{\delta k})=\xi\hbar v_x \delta k_x+i\hbar v_y\delta k_y,
\label{eqfk dirac}
\ee
where the velocities along the $x$ and $y$ directions are
\be
v_x= \sqrt{\frac{4-\lambda^{2}}{3}}v_F,\hspace{0.2cm} v_y= \lambda v_F, 
\label{vxy}
\ee
with $v_F=\frac{3at}{2\hbar}$ being the Fermi velocity in the undeformed lattice.
As $\lambda$ approaches $2$, the linear term in $\delta k_x$ vanishes, so it becomes necessary to include a quadratic term in $\delta k_x$ \cite{Bahat}.
For $\lambda>2$, the velocity $v_x$ becomes imaginary, and the linearized Hamiltonian is no longer valid.
To describe the system in this regime, we expand the function $f_{\lambda}(\vec{k})$ around the $M$ point by writing $k_x=\delta k_x+k_M$, $k_y=\delta k_y$, yielding 
\be
f_{\lambda}(\vec{k})=\Delta +\frac{\hbar^2k_x^2}{2m}+i\hbar v_yk_y,
\label{eqfk univ}
\ee
where $k_M=\frac{\xi m v_x}{\hbar}$ is the position of the Dirac point $D_\xi$ relative to the $M$ point, and $\Delta=-\frac{1}{2}mv_x^2$ defines the gap parameter. 
The effective mass $m$, is obtained by expanding $f_{\lambda}(k_x,k_y=0)$ (Eq. (\ref{fk})) near the $M$ point to first order in $k_x$, giving
\be
\Delta=(\lambda-2)t, \hspace{1cm} m=\frac{8\hbar^2}{3(2+\lambda)a^2t}.
\ee

The Hamiltonian (\ref{eq1}) with $f_{\lambda}(\vec{k})$ given by Eq. (\ref{eqfk univ}) describes the physics of the system for all values $\lambda\geq 1$ and can be written as
\be
H=\left(\Delta +\frac{\hbar^2k_x^2}{2m}\right) S^{\varphi}_x+\hbar v_yk_y S^{\varphi}_y,
\label{unv_hamlt}
\ee
where the pseudospin matrices are
 \be
\begin{split}
 S_{x}^{\varphi}=\begin{pmatrix}
0 &\cos\varphi  &0 \\ 
\cos\varphi & 0 &\sin\varphi  \\ 
 0& \sin\varphi &0 
\end{pmatrix},&\\  
S_{y}^{\varphi}=i \begin{pmatrix}
0 &-\cos\varphi  &0 \\ 
\cos\varphi & 0 &-\sin\varphi  \\ 
 0& \sin\varphi &0 
\end{pmatrix}.
\end{split}
\label{pseudospinxy} 
\ee
It is worth noting that this Hamiltonian reproduces the universal Hamiltonian derived by G. Montambaux \textit{et al.} \cite{Gilles1} in the case of graphene ($\alpha=0$), highlighting its role as a unified description for the deformed $\alpha-T_3$ lattice across all deformation strengths.

\begin{figure*}[t]
  \centering
\setlength{\unitlength}{1mm}
\includegraphics[width=1\textwidth]{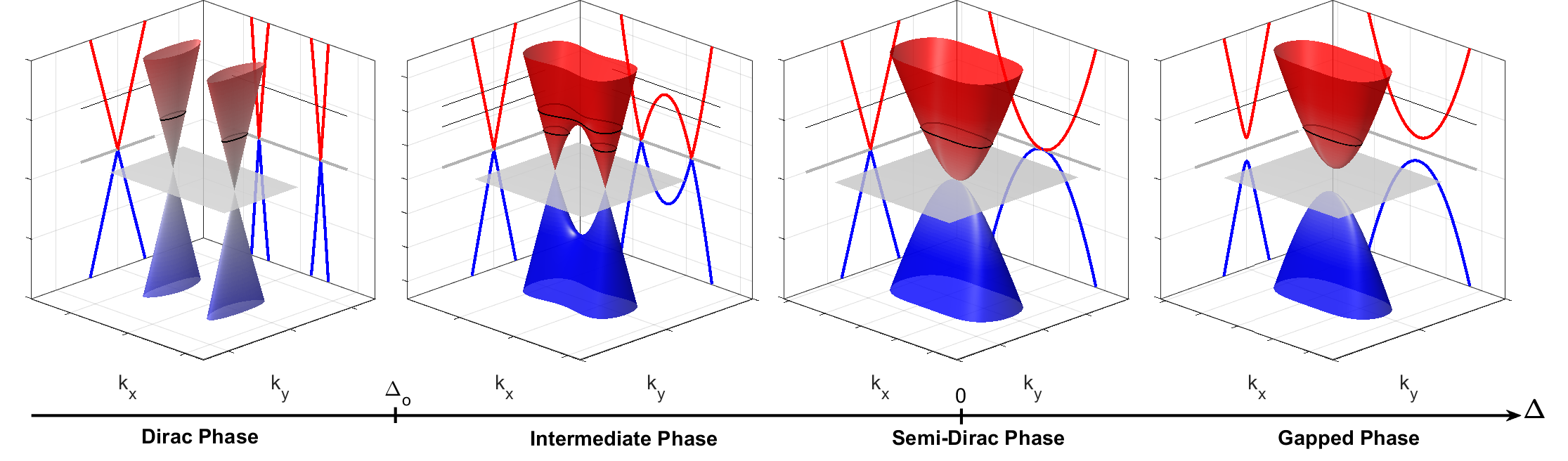}
\caption{(Color online) Energy spectrum of the deformed $\alpha-T_3$ model for different values of $\Delta$. The black lines represent the particle energy $E$.  Dirac phase: $\Delta<\Delta_o<0$ and $E<<\left |\Delta  \right |$, where the spectrum consists of two anisotropic Dirac cones and a flat band. Intermediate phase: $\Delta_o <\Delta \leq 0$ with $E\sim\left |\Delta  \right |$, characterized by the coexistence of massless and massive particles along the $x$ direction. Semi-Dirac phase: $E > \Delta = 0$. The dispersion is linear along the $k_y$ direction (massless particles) and quadratic along the $k_x$  direction (massive particles), with a flat band at $E=0$. 
 Gapped phase: $E > \Delta > 0$, where the spectrum is fully gapped, with a flat band at $E=0$ and the particles are massive. 
In all cases with $\Delta \leq 0$, particles remain massless along the $y$ direction. }
\label{figure2}
\end{figure*}

In the low-energy limit, variations in $\Delta$ (or equivalently in $\lambda$) give rise to three distinct phases, as illustrated in Fig. \ref{figure2}. For $\Delta<\Delta_o<0$ and $E<<\left |\Delta  \right |$, the spectrum consists of two separate Dirac cones located at $D_\pm$ along with a flat band, defining the Dirac phase \cite{Adroguer}. In this phase, the Dirac cones are anisotropic, with different velocities along the $x$ and $y$ directions [Eq. (\ref{vxy})], and the particles behave as massless Dirac fermions.
For $\Delta_o<\Delta \leq 0$ and energies $E\sim\left |\Delta  \right |$, the system enters an intermediate phase, characterized by the coexistence of massless and massive particles along the $x$ direction. 
The gapped phase appears for $\Delta>0$, in which the particles are massive. In particular, at $\Delta=0$, the band structure is linear along $k_y$ direction and parabolic along $k_x$ direction, with a flat band at $E=0$, corresponding to the $\alpha-T_3$ semi-Dirac model \cite{Piechon}. Notably, for $\Delta \leq 0$, the dispersion remains linear along the $y$ direction, so particles behave as massless Dirac fermions along that axis. Finally, the flat band is robust against deformation.

Graphene can sustain large elastic deformations, with a critical strain amplitude of approximately $\varepsilon \approx 0.2$, as predicted theoretically \cite{Pereira} and confirmed experimentally \cite{Kim2009}. Following the tight-binding approach of Ref. \cite{Pereira}, the strain-dependent coupling parameter $\lambda$ is given by $\lambda = e^{\eta \varepsilon}$, where $\eta$ depends on the direction of the applied strain and reaches its maximum value of $\eta \approx 2.95$ for uniaxial deformation along the zigzag direction. In this regime, values of $1 \leq \lambda \leq 1.8$ correspond to experimentally accessible strain amplitudes in graphene. Nevertheless, the merging of Dirac points can be more readily realized in artificial graphene systems, where larger deformations or tunable lattice parameters can be engineered.


\subsection{Topological properties }

We next explore how deformation affects the system's topological behavior, with particular attention to its influence on the Berry phase, Berry connection, and Berry curvature.
The Berry connection $\mathcal{\bm A}_\lambda^s(\bm k)=i \left \langle \Psi_\lambda^s(\bm k)|\bm \nabla_{\bm k}\Psi_\lambda^s(\bm k) \right \rangle$ of each band $E_s$ is given by 
\be
\mathcal{\bm A}_\lambda^\pm(\bm k)=\frac{1}{2} \cos2\varphi \nabla_{\bm k}\theta_{\lambda},\hspace{1cm} \mathcal{\bm A}_\lambda^0(\bm k)=-2 \mathcal{\bm A}_\lambda^\pm(\bm k)
\ee 
where the phase is defined as $\theta_{\lambda} =\arg (f_{\lambda}(\bm k))$, with $f_{\lambda}(\bm k)$ is given by Eq. (\ref{eqfk univ}).
More explicitly, the quantity $\nabla_{\bm k}\theta_{\lambda}$ can be written in Cartesian coordinates as
\be
\nabla_{\bm k}\theta_{\lambda}=\left(-\frac{dK_x}{dk_x}\frac{K_y}{K_x^2+K_y^2,},\frac{dK_y}{dk_y}\frac{K_x}{K_x^2+K_y^2},0\right),
\ee
where $K_x=\Delta +\frac{\hbar^2k_x^2}{2m}$ and $K_y=\hbar v_yk_y$. Note that, for $1\leq\lambda\leq2$, $\nabla_{\bm k}\theta_{\lambda}$ is not defined at the Dirac points $D_\pm(\pm k_d,0)$ with $k_d=\sqrt{-\frac{2m\Delta}{\hbar^2}}$. 
From the Berry connection we obtain, for each band, the Berry curvature $\bm \Omega_\lambda^s(\bm k)=\nabla_{\bm k}\times\mathcal{\bm A}_\lambda^s(\bm k)=\Omega_\lambda^s(\bm k) \hat{\bm k} $ and the Berry phase $\Phi_B^s= \oint_\mathscr{C} \mathcal{\bm A}_\lambda^s(\bm k).d{\bm k}=\int_\mathcal{S}  \Omega_\lambda^s(\bm k) d^2\bm k $, where $\mathcal{S}$ is the surface enclosed by the constant energy contour $\mathscr{C}$. 
Using the polar coordinates (i.e., $(K_x,K_y)\rightarrow(K,\theta_{\lambda})$), as in Ref. \cite{Iurov2019}, the Berry curvature for the conical bands, is obtained as
\be
\Omega_\lambda^\pm(\bm k)=\pi \cos2\varphi \left[\delta(k_x-k_d)-\delta(k_x+k_d)\right]\delta(k_y).
\ee

Consequently, the Berry phase in this system is topological rather than geometrical, meaning that it does not depend on the specific shape of the closed contour $\mathscr{C}$ as long as it encloses the Dirac point.
It is given by
\be 
 \Phi_B^\pm=\pi \cos2\varphi W_\mathscr{C},
\ee 
where $W_\mathscr{C}=\oint_\mathscr{C} \nabla_{\bm k}\theta_{\lambda}.d{\bm k}/2\pi$ is the winding number, which acts as a topological invariant.
Before the merging of the Dirac points ($1 \le \lambda < 2$), the system hosts two separate Dirac cones, denoted $D_+$ and $D_-$, each characterized by a quantized topological charge: $W_{\mathscr{C}} = +1$ and $W_{\mathscr{C}} = -1$ respectively. Physically, this non-zero winding number protects the Dirac quasiparticles against lattice deformation, ensuring their robustness.
After the merging transition, however, the two opposite topological charges annihilate each other, resulting in $W_{\mathscr{C}} = 0$. This corresponds to a Lifshitz-type topological phase transition \cite{Lifshitz} involving a change in the topology of the Fermi surface: initially composed of two discrete points (before merging), then a single touching point at the critical value, and finally no closed Fermi contour after the transition.

Note that throughout this transition, time-reversal symmetry is preserved. Consequently, the Chern number $C=\int_{BZ} \Omega_\lambda^s(\bm k) d^2\bm k/2\pi=0$.
Thus, although the system exhibits a topological transition via the annihilation of opposite winding-number defects, it does not become a Chern insulator, since no symmetry is broken to permit a non-zero net Berry curvature.
In contrast, it has been shown that the $\alpha-T_3$ illuminated by circularly polarized radiation undergoes a topological transition from a semimetal to a Chern insulator with a nonzero Chern number due to the breaking of time-reversal symmetry \cite{Dey2019}.

\section{Effect of Deformation on Tunneling Across a Sharp $\mathit{np}$ Junction} \label{III}

In this section, we investigate the effect of deformation (illustrated in Fig. \ref{figure2}) on tunneling through an $\mathit{np}$ junction. We focus on the case where the junction is oriented parallel to the deformation direction, where particles evolve from massless to massive Dirac fermions as the deformation increases. 
In contrast, when the junction is oriented perpendicular to the deformation direction, the particles remain massless Dirac fermions regardless of the deformation strength. Since this situation does not qualitatively modify the tunneling behavior, it is discussed separately in Appendix \ref{KT y}.

For the parallel configuration, the Hamiltonian is given by Eq. (\ref{unv_hamlt}) with the potential 
\be 
 V(x)=V_o \Theta(x),  
\label{pot x}
\ee
where $\Theta(x)$ denotes the Heaviside step function.\\
The potential step is sharp but is assumed to vary over a length scale larger than the in-plane interatomic distance and smaller than the electron wavelength, ensuring that it does not induce intervalley scattering.
It is uniform along the $y$ direction, so the parallel component of the wave vector, $k_y$, is conserved.
Accordingly, the wave function can be expressed as $\Psi \left ( x,y \right )=\psi \left (x \right )e^{ik_{y}y}$.

For an energy $E = V_o/2$, transverse momentum $k_y$, and gap parameter $\Delta$, there exist four possible longitudinal wave vectors $k_x^{ss'}$ satisfying the equation $E^2 = |f_{\lambda}(k_x, k_y)|^2$, which can be expressed as
\be
k_{x}^{ss'}=s\sqrt{\frac{2m}{\hbar^2}}\sqrt{-\Delta+s' \sqrt{E^2-\hbar^2v_y^2k_{y}^2}},
\label{kx}
\ee
where $s,s'=\pm 1$ label the four solutions, and $f_{\lambda}(k_{x},k_{y})$ is given in Eq. (\ref{eqfk univ}).
This wave vector may be real or imaginary, corresponding respectively to a propagating or an evanescent mode.
The corresponding normalized eigenstate is of the form
\be 
\psi_{\nu}^{(s,s')}(x)= \frac{1}{\sqrt{2}} \begin{pmatrix}
\cos\varphi e^{-i\theta_{s'}} \\ 
\nu\\ 
\sin\varphi e^{i\theta_{s'}} 
\end{pmatrix}e^{ik_{x}^{ss'}x},
\label{psix np}
\ee 
where $\nu=\pm 1$ is the band index and  

\be 
\theta_{s'}=\arg \left [s'\sqrt{E^2-\hbar^2v_y^2k_{y}^2}+i\hbar v_y k_{y}  \right ], 
\label{angles}
\ee
which fixes the in-plane pseudospin orientation relative to the $k_x$ axis.
The pseudospin expectation values are
\be
\begin{aligned}
&\left \langle S_x \right \rangle_{\nu}^{s'}=\expval{S_{x}^{\varphi}}{\psi_{\nu}^{s'}}=\nu \cos \theta_{s'}, \\
&\left \langle S_y \right \rangle_{\nu}^{s'}=\expval{S_{y}^{\varphi}}{\psi_{\nu}^{s'}}=\nu \sin \theta_{s'}. 
\end{aligned}
\label{spin direction}
\ee

\begin{figure}[h!]
  \centering
\setlength{\unitlength}{1mm}
\includegraphics[width=0.48\textwidth]{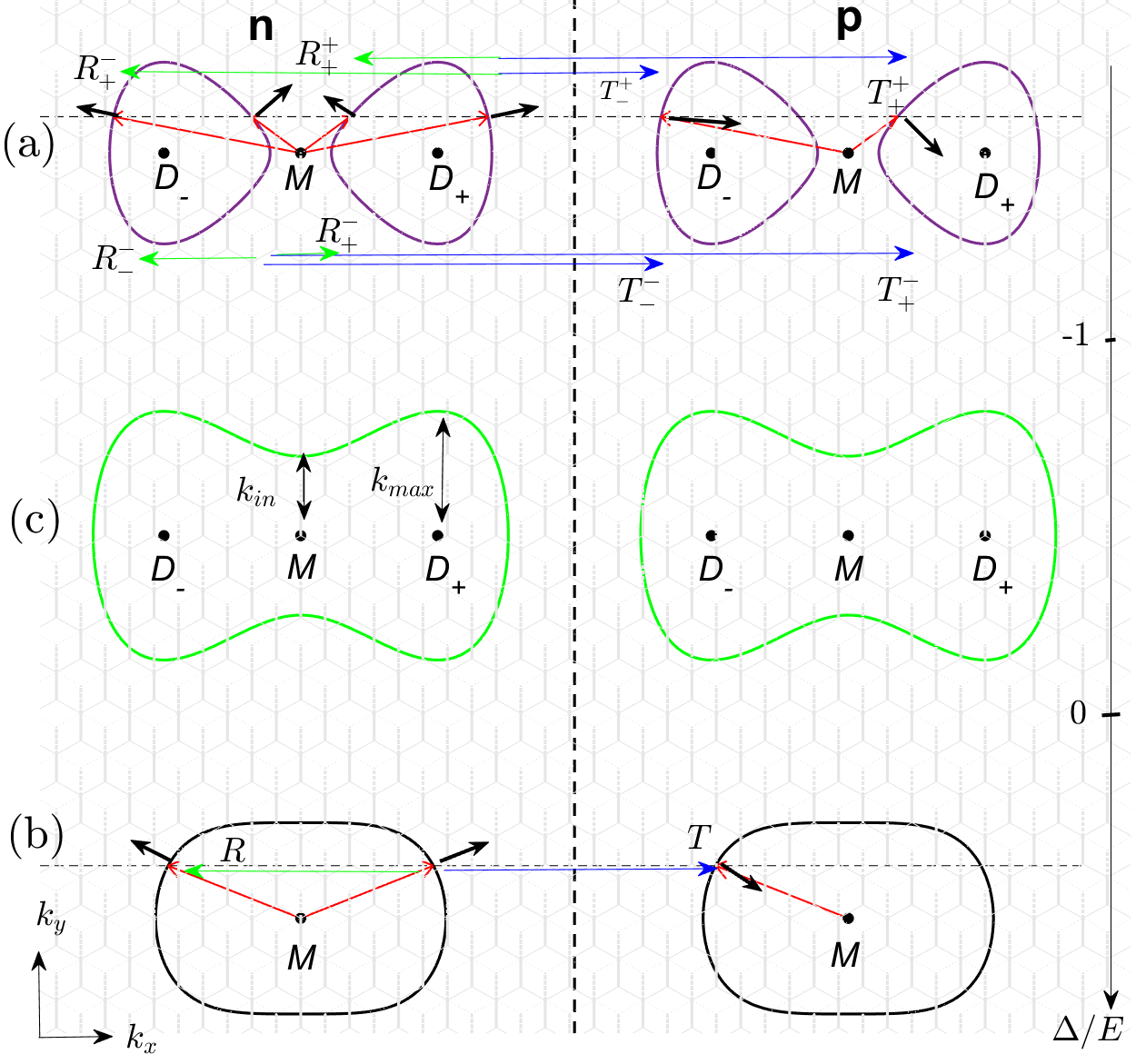}
\caption{(Color online) Schematic of transmission and reflection probabilities for an $np$ junction along the deformation axis ($y$ direction) at $E = V_0/2$.
(a) For $\Delta/E<-1$, the Fermi surfaces are disconnected. For a given parallel wave vector $k_y$ (horizontal dashed line), there are four propagating modes (red arrows) corresponding to four transmission and four reflection probabilities. The black arrows indicate the group velocities, showing the direction of propagation.
(b) For $0<\Delta/E<1$, the Fermi surfaces are connected. For a given $k_y$, there is a single transmission and a single reflection probability.
(c) For $-1 < \Delta/E < 0$, the Fermi surfaces are partially connected. For $k_{\mathrm{in}} < |k_y| < k_{\mathrm{max}}$, there are four transmission and four reflection probabilities as in (a), while for $|k_y| < k_{\mathrm{in}}$, there is only one transmission and one reflection probability, as in (b).}    
\label{figtrans}
\end{figure}

The tunneling across the $\mathit{np}$ junction is determined by the deformation of the lattice, characterized by the gap parameter $\Delta$. 
Depending on the ratio $\Delta/E$, three distinct situations arise, as shown in Fig. \ref{figtrans}.

(i) \textit{Disconnected Fermi surfaces} ($\Delta/E<-1$):\\ In this situation, the two Fermi surfaces are completely disconnected. The four longitudinal wave vectors $k_{x}^{ss'}$ are real when $\left|k_{y} \right|<k_{max}=\frac{E}{\hbar v_y}$, leading to four transmission channels [Fig. \ref{figtrans} (a)]. The wave functions in the  $n$ and $p$ regions are 
\be
\begin{aligned}
&\psi(x<0)=\psi_{n}^{(s,s)}(x)+r_{s}^{s}\psi_{n}^{(s, -s)}(x)+r_{-s}^{s}\psi_{n}^{(-s, s)}(x), \\
&\psi(x>0)=t_{s}^{s}\psi_{p}^{(s, -s)}(x)+t_{-s}^{s}\psi_{p}^{(-s, s)}(x), 
\end{aligned}
\label{psi4 np}
\ee 
where $t_{s}^{s}$ ($t_{-s}^{s}$) is the intravalley (intervalley) transmission amplitude, and $r_{s}^{s}$ ($r_{-s}^{s}$) is the corresponding reflection amplitude. The wave function $\psi_{n(p)}^{(s s')}(x)$ in  $n$ ($p$) region is given by Eq. (\ref{psix np}) with $\nu=1$ ($\nu=-1$).

(ii) \textit{Connected Fermi surfaces} ($0<\Delta/E<1$):\\ In the gapped phase, the Fermi surfaces are connected [Fig. \ref{figtrans} (b)]. The longitudinal momentum $k_{x}^{s+}$ is real when $\left | k_{y} \right |<k_{in}=\frac{\sqrt{E^2-\Delta^2}}{\hbar v_y}$, while $k_{x}^{s-}$ becomes imaginary, resulting in one transmission channel. The total wave functions now include both propagating and evanescent components
\be
\begin{aligned}
&\psi(x<0)=\psi_{n}^{(+,+)}(x)+r\psi_{n}^{(-,+)}(x)+A\psi_{n}^{(-,-)}(x), \\
&\psi(x>0)=t\psi_{p}^{(-,+)}(x)+B\psi_{p}^{(+, -)}(x), 
\end{aligned}
\label{psi2 np}
\ee 
where $r$ and $t$ are the reflection and transmission amplitudes, and $A$ and $B$ are the amplitudes of the evanescent modes.

(iii) \textit{Partially connected Fermi surfaces} ($-1 < \Delta/E < 0$):\\
In this case [Fig. \ref{figtrans} (c)],  the behavior depends on the value of $k_y$: 
for $k_{in}<\left | k_{y} \right |<k_{max}$, there are four transmission channels, with the wave function given by Eq. (\ref{psi4 np}) as in situation (i);
for $\left| k_{y} \right| < k_{\mathrm{in}}$, the Fermi surfaces are connected as in case (ii), yielding a single transmission channel, with the wave functions given by Eq. (\ref{psi2 np}).

For electronic systems, a more directly measurable quantity is the ballistic conductance, which is given by using the Landauer-B\"uttiker formula \cite{Blanter}
\be
G_x=2G_o\sum_{k_{y}}T(k_{y}),
\label{conductancex dirac}
\ee
where $T$ is the total transmission probability, $G_o=\frac{e^2}{h}$ and the factor $2$ accounts for the spin degeneracy.
All possible transmission and reflection probabilities, along with the corresponding conductance expressions, are provided in Appendix \ref{A}.

We plot the conductance $G_x$ [Fig. \ref{allnp} (a)] and the transmission probability at normal incidence [Fig. \ref{allnp} (b)] as a function of $\Delta/E$ for various values of the parameter $\alpha$.
The conductance $G_x$ is normalized to the conductance $G_x^{SKT}$, the conductance in the SKT regime of the Dirac phase, given by Eq. (\ref{condskt}).
Both KT and SKT occur when $\Delta<\Delta_o$, corresponding to the Dirac phase, where particles behave as massless Dirac fermions.
The tunneling properties in this phase are analyzed for various junction orientations $\beta$ with respect to the deformation axis in Appendix \ref{DP}.
KT occurs for any value of $\alpha$ and $\beta$; perfect transmission is achieved at normal incidence only for $\beta = 0$ and $\beta = \pi/2$, 
while for other $\beta$ values it appears at oblique incidence. This behavior arises from the conservation of pseudospin between the incident and transmitted waves.
In the dice lattice ($\alpha=1$), however, SKT effect occurs for any $\beta$, an effect associated with the conservation of the transverse pseudospin component.
We conclude that the electronic transport properties in the Dirac phase of the deformed $\alpha-T_3$ lattice remain similar to those in the undeformed case, a consequence of the massless Dirac fermion nature of the carriers in this phase.
Finally, the KT and the SKT disappear for $\Delta>\Delta_o$ (see Fig. \ref{allnp}), marking the transition from the Dirac to the intermediate phase.

In the gapped phase ($0<\Delta/E<1$), the two Fermi surfaces merge, leading to a quadratic dispersion relation along the $x-$direction, indicating that particles traversing the junction are massive.
At normal incidence, perfect reflection occurs for all $\alpha$, a phenomenon known as AKT, as shown in Fig. \ref{allnp}(b).
For the dice lattice ($\alpha=1$), the transmission probability vanishes for all incident angles [Fig. \ref{allnp}(a)], despite the presence of states at the junction interface.
This effect, referred to as ASKT \cite{Yonatan}, is the counterpart of SKT. 
These phenomena (AKT and ASKT) arise from pseudospin conservation and can be understood as follows.
From Eq. (\ref{psi2 np}), the incident, reflected, and transmitted progressive waves share the same pseudospin angle $\theta_+$, so that
the pseudospin of the incident wave is aligned with that of the reflected wave but opposite to the transmitted wave, $\bm S_i=\bm S_r=-\bm S_t=(\cos \theta_+,\sin \theta_+)$. 
In contrast, the evanescent waves are characterized by a different pseudospin angle $\theta_-$.
On the other hand, the condition for a perfect reflection is satisfied when the transmission amplitude, given in Eq. (\ref{t npii}), vanishes. This condition leads to the following relations
\be
\begin{aligned}
&\cos\theta_+\pm \cos\theta_-=0, \\ 
&\cos 2\varphi \left(\sin\theta_+\pm \sin\theta_-\right)=0. 
\end{aligned}
\ee 
The first condition reflects the conservation of the transverse pseudospin component between the propagating and evanescent modes, which holds in our case (but not when $E \neq V_0/2$). The second condition is fulfilled either for the dice lattice ($\alpha=1$) at any incidence angle, leading to ASKT, or for normal incidence for all $\alpha$, leading to AKT.
 This demonstrates that the conservation of pseudospin between the incident and reflected waves, as well as between the progressive and evanescent waves, is the underlying mechanism responsible for AKT, while conservation of the transverse components of the pseudospin between these waves gives rise to ASKT.

In the intermediate phase ($\Delta_o <\Delta \leq 0$), two sub-phases can be identified:

 (i) for $\Delta/E<-1$, the Fermi surfaces are disconnected (see Fig. \ref{figtrans} (a)) but the dispersion relation around the Dirac points $D_{\pm}$ is no longer linear.
This indicates that the current along the $x-$direction consists of a mixture of massive and massless carriers.
In this sub-phase, as the two Fermi surfaces approach each other, intervalley reflections $R_{-s}^{s}$ [Eq. (\ref{T np})] emerge, and the resulting intervalley scattering leads to the disappearance of KT and SKT.

 (ii) For $-1<\Delta/E<0$, the Fermi surfaces are partially connected (see Fig. \ref{figtrans} (c)).
 Perfect reflection occurs when the Fermi surfaces are connected ( $\left| k_{y} \right| < k_{\mathrm{in}}$), particularly at normal incidence, giving rise to the AKT effect.
When the Fermi surfaces are disconnected ($k_{in}<\left | k_{y} \right |<k_{max}$), perfect reflection is suppressed, as in sub-phase (i), preventing the emergence of the ASKT regime.

However, we observe that the conductance exhibits two opposite trends as a function of the parameter $\alpha$ when the gap parameter $\Delta$ increases [Fig. \ref{allnp} (a)]. 
Firstly, the junction becomes more transparent with increasing $\alpha$, indicating that massless carriers dominate the current along the $x-$direction.
 Secondly, the junction becomes less transparent as $\alpha$ increases, reflecting the predominance of massive carriers.

Finally, a remarkable result is the observation of a transition from KT to AKT for all values of the parameter $\alpha$, similar to the case of the deformed HCL ($\alpha=0$) \cite{Bahat}. Likewise, a transition from SKT to ASKT occurs in the deformed dice lattice ($\alpha=1$) for an energy $E=V_o/2$.

\begin{figure}[h!]
  \centering
\setlength{\unitlength}{1mm}
\includegraphics[width=0.5\textwidth]{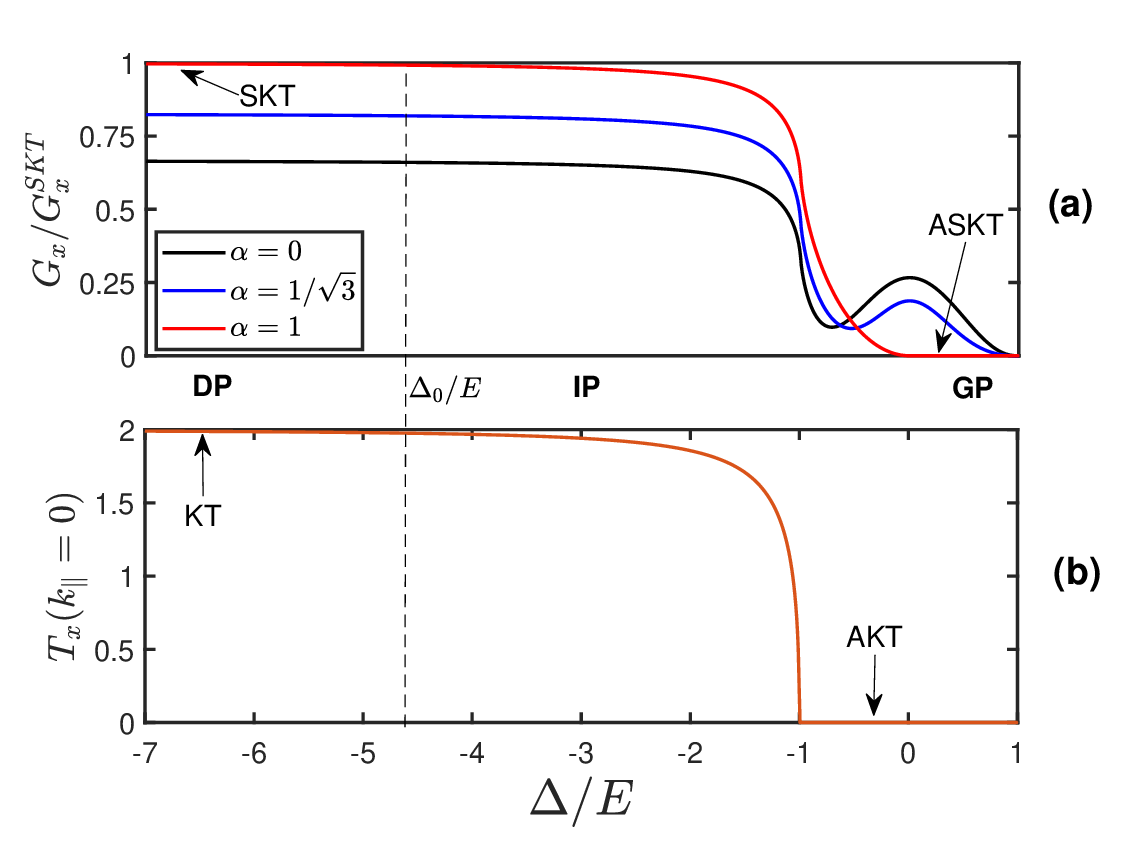} 
\caption{(Color online) (a) The conductance in units of $G_x^{SKT}$ as a function of $\Delta/E$ and for the three values of $\alpha$. (b) Transmission probability at normal incidence, independent of $\alpha$. 
The three phases (DP, IP, GP, as in Fig. \ref{figure2}) are indicated.
Here, $G_x^{SKT}$ is the conductance of the SKT regime in the Dirac phase [Eq. (\ref{condskt})] and the step height is set to $V_o=0.08 t$.  }
\label{allnp}
\end{figure}

We now turn to the role of the flat band in the tunneling properties. We emphasize that, in our calculations, the tunneling process involves only the conduction and valence bands; the flat band does not participate directly. Moreover, the flat band remains robust against the applied deformation. We have shown that the tunneling characteristics depend explicitly on the parameter $\alpha$, which allows us to compare the cases with ($\alpha \neq 0$) and without ($\alpha=0$) a flat band contribution. In particular, the junction becomes more transparent when a flat band is present in the spectrum ($\alpha=0 \rightarrow \alpha \neq 0$) for massless carriers, whereas it becomes less transparent when the carriers are massive.
The addition of a mass term in the Hamiltonian of the undeformed $\alpha-T_3$ model opens an energy band gap with a flat band lying inside the gap at a tunable energy position \cite{Piechon,Betancur}. It has been shown in Ref. \cite{Betancur} that the tunneling through a junction depends sensitively on the position of this flat band, even though it does not participate directly in the tunneling process, as is the case in our calculations.

An interesting perspective would be to explore situations where the flat band acquires a finite dispersion, for example when the $\alpha-T_3$ material
is subjected to nonresonant irradiation by circularly polarized light \cite{Iurovprb2023,Dey2018}. In this case, the flat band may develop additional propagating modes that contribute to transport.

\section{Effect of the junction orientation on the Klein tunneling} \label{IV}

\begin{figure}[h!]
  \centering
\setlength{\unitlength}{1mm}
\includegraphics[width=0.48\textwidth]{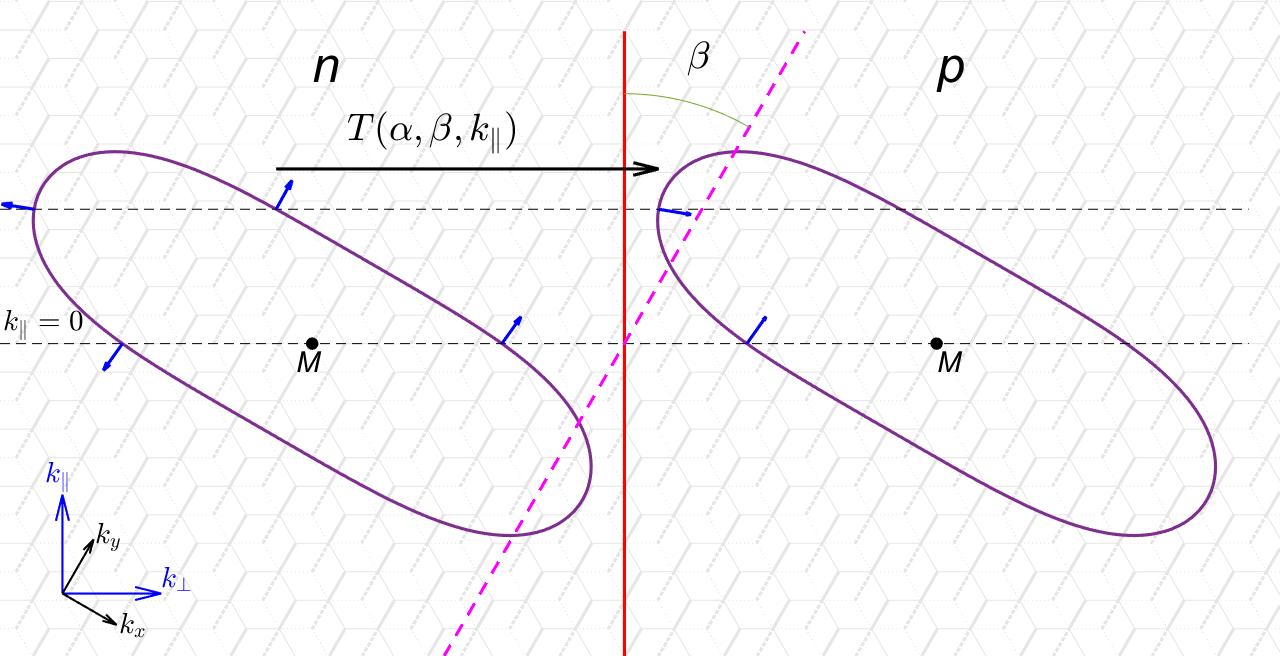} 
\caption{(Color online) Schematic representation of the transmission probability in the semi-Dirac phase across an $np$ junction oriented at an angle $\beta$ relative to the deformation axis (magenta dashed line). Blue arrows indicate the group velocities, and the horizontal dashed lines show the conserved parallel wave vector, $k_\parallel$.  }
\label{fig61}
\end{figure}

We study tunneling across the $\mathit{np}$ junction as a function of its orientation angle $\beta$ relative to the deformation axis in the semi-Dirac phase ($\Delta=0$) (see Fig. \ref{fig61}).
The Hamiltonian is $H_\beta=H+V(x_\perp,x_\parallel)$,
where $H$ is given in Eq. (\ref{unv_hamlt}), with the momentum components transformed as
\be
\begin{aligned}
&k_x=\cos\beta k_{\perp}-\sin\beta k_{\parallel}\\
&k_y=\sin\beta k_{\perp}+\cos\beta k_{\parallel}
\end{aligned}
\label{convert k}
\ee
with $k_\perp$ and $k_\parallel$ denoting the transverse and longitudinal components to the junction, respectively.
The potential is 
\be 
 V(x_\perp,x_\parallel)=V_o \Theta(x_\perp),  
\label{pot xperp}
\ee
where $x_\perp$ and $x_\parallel$ are the directions perpendicular and parallel to the junction, respectively, and $\Theta(x_\perp)$ is the Heaviside step function.
Since the potential step is translationally invariant along $x_\parallel$, the parallel component of the wave vector, $k_\parallel$, is conserved. Consequently, the wave function can be written as $\Psi \left ( x_\perp,x_\parallel \right )=\psi \left ( x_\perp \right )e^{ik_{\parallel}x_\parallel}$.

For an energy $E=V_o/2$, a parallel momentum $k_{\parallel}$ and an orientation angle $\beta \neq 0,\pi/2$, the wave functions in the  $n$ and $p$  regions are 
\be
\begin{aligned}
\psi(x_\perp<0)&=\psi_{n}^{(-s,1)}(x_\perp)+\rho \psi_{n}^{(-s,-1)}(x_\perp)\\&+A\psi_{n}^{(s,-1)}(x_\perp), \\
\psi(x_\perp>0)&=\tau \psi_{p}^{(-s,-1)}(x_\perp)+B\psi_{p}^{(s,1)}(x_\perp), 
\end{aligned}
\label{psi2 np sd}
\ee 
where $s=\mathrm{sign}(k_\parallel)$, $\rho$ and $\tau$ are the reflection and transmission amplitudes, and $A$ and $B$ denote the amplitudes of evanescent waves.
The wave functions $\psi_{n(p)}^{(s,s')}(x_\perp)$ are derived in Appendix \ref{C} (see Eq. (\ref{psi sd})).
Applying the matching condition (Eq.~\ref{match cond gen}), we numerically obtain the amplitudes $\rho$, $\tau$, $A$, and $B$, from which the transmission probability is calculated as
\be
T(\alpha,\beta,k_\parallel)=\frac{J_\perp\left[\psi^{(-s,-1)}_p\right]}{J_\perp\left[\psi^{(-s,1)}_n\right]}|\tau|^2,
\ee
where $J_\perp$ is the transverse probability current (Eq.~\ref{trans curr}).
In the limiting cases, $\beta=0$ and $\beta=\pi/2$, the above expression reduces to the transmission probabilities derived in Section \ref{III} (see Eq. (\ref{T(ii)}) and in Appendix \ref{KT y} (see Eq. (\ref{T dirac y})) by taking $\Delta=0$.
In particular, for $k_\parallel = 0$, an analytic expression for the transmission probability can be derived
\be
T(\alpha,\beta,k_\parallel=0)=\frac{4\sin^2\beta \cos^2 2\varphi}{4\sin^2\beta \cos^2 2\varphi+k_o^2},
\label{Tkpar0}
\ee
where $k_o=\sqrt{-2\sin^2\beta+\sqrt{4\sin^4\beta+\varepsilon^2\cos^4\beta/9}}$, with the dimensionless energy $\varepsilon = E/t$.

Since we have the transmission probability, we can deduce the conductance using the Landauer-B\"uttiker formula \cite{Blanter} 
\be
G_\beta=2G_o\sum_{k_{\parallel}}T(\alpha,\beta,k_\parallel),
\label{conductance sdirac}
\ee
where $G_o=\frac{e^2}{h}$ and the factor $2$ accounts for the spin degeneracy.

\begin{figure}[h!]
  \centering
\setlength{\unitlength}{1mm}
\includegraphics[width=0.5\textwidth]{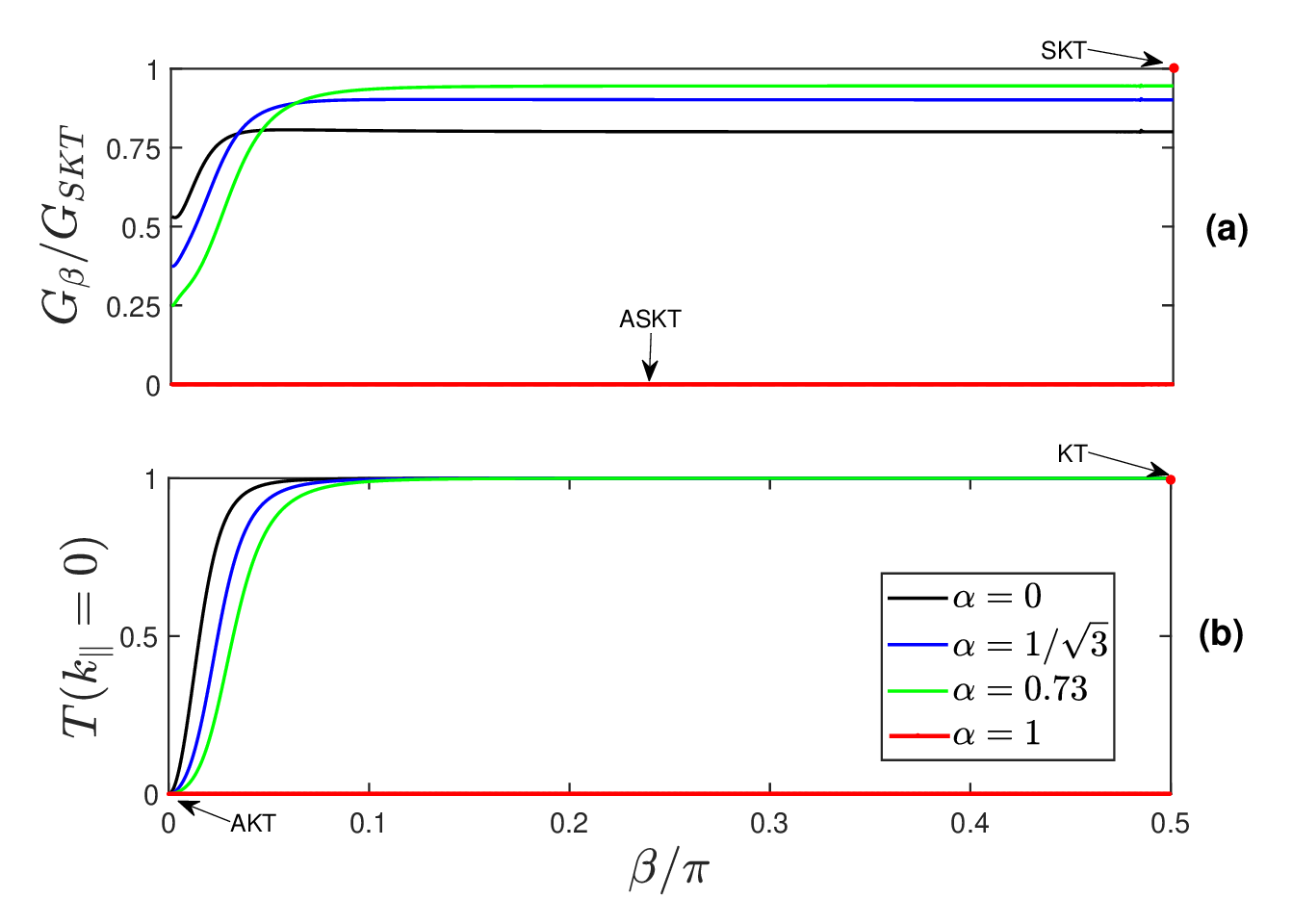}
\caption{(Color online) 
(a) Conductance, in units of $G_{SKT}$, as a function of $\beta/\pi$ for different values of $\alpha$. 
(b) Transmission probability as a function of $\beta$ for $k_\parallel = 0$,
 The potential step has a height of $V_o=0.08 t$.}
\label{plotsdirac}
\end{figure}

Figure \ref{plotsdirac}(b) shows the transmission probability as a function of $\beta$ for $k_\parallel = 0$, corresponding to normal incidence at $\beta = 0$ and $\beta = \pi/2$. A clear transition is observed from AKT at $\beta = 0$ to KT at $\beta = \pi/2$ for all values of the particle parameter $\alpha$, which determines whether the particles are massive ($\beta = 0$) or massless ($\beta = \pi/2$). 
In Fig. \ref{plotsdirac}(a), the conductance $G_\beta / G_{SKT}$ is shown as a function of $\beta$ for different values of the parameter $\alpha$. Here, $G_{SKT}$ denotes the conductance in the SKT regime, given by Eq. (\ref{conductance sdirac}) with $T(\alpha, \beta, k_\parallel) = 1$, which reads
\be 
G_{SKT}=2G_oL_yk_{max}/\pi,
\ee
where $k_{max}=\sqrt{2mE/\hbar^2}$. 
Notably, for the dice lattice ($\alpha=1$), an abrupt transition  from AKT to KT and from ASKT to SKT occurs at $\beta=\pi/2$. 
This behavior originates from the presence of evanescent modes (corresponding to massive particles) in the range $0 \leq \beta<\pi/2$ (see Eq. (\ref{psi2 np sd})), which suppress transmission until $\beta$ reaches $\pi/2$, where perfect tunneling is restored.

The conductance exhibits two distinct behaviors as $\beta$ increases, as depicted in Fig.~\ref{plotsdirac}(a). First, for increasing $\alpha$, the junction becomes less transparent, and the particles traversing it are predominantly massive. Second, for $\alpha \neq 1$, the conductance shows an opposite trend, reflecting the dominance of massless particles.

\section{Conclusions} \label{V}
In this work, we have investigated the effect of a uniaxial deformation on tunneling across an $\mathit{np}$ junction within the $\alpha-T_3$ model, focusing on the case where the particle energy equals half of the step height.
We have shown that, under compressive deformation, the Dirac cones move toward each other and eventually merge, while the flat band remains unaffected.

In a first step, we have shown that a uniaxial deformation applied along the junction direction induces a transition in the tunneling regime across a sharp $\mathit{np}$ interface. Specifically, the system evolves from KT to AKT for all values of the parameter $\alpha$. 
 In the dice lattice ($\alpha=1$), the deformation further drives a transition from SKT to ASKT. These results highlight the fundamental change in the quasiparticle nature, from massless Dirac fermions in the undeformed phase to massive Dirac fermions in the gapped phase.  
For moderate deformation (within the Dirac phase) and for arbitrary junction orientation, we find that KT persists for all values of $\alpha$, as a direct consequence of pseudospin conservation. In the dice lattice ($\alpha=1$), SKT is recovered, as in the pristine case, reflecting the conservation of the longitudinal pseudospin component. Moreover, the junction becomes more transparent as $\alpha$ increases, consistent with earlier results for the undeformed $\alpha$–$T_3$ lattice. 
In contrast, when the Fermi surfaces merge beyond a critical deformation (the gapped phase), AKT arises for all $\alpha$, while ASKT is obtained for $\alpha=1$.
The AKT effect originates from pseudospin conservation involving both propagating and evanescent modes, whereas ASKT results from the conservation of the transverse pseudospin components. In this regime, the junction becomes less transparent as $\alpha$ increases.
When the junction is oriented perpendicular to the deformation direction, the particles remain massless Dirac fermions regardless of the deformation, so both KT and SKT are unaffected.

In a second step, we studied the effect of the rotation of the $\mathit{np}$ junction on tunneling in the semi-Dirac phase. A transition from KT to AKT occurs for all values of $\alpha$, reflecting the change from massless Dirac fermions when the junction is perpendicular to the deformation direction to massive Dirac fermions when it is parallel. Notably, for the dice lattice ($\alpha=1$), an abrupt transition from AKT to KT and from ASKT to SKT is observed. This behavior is due to the presence of evanescent modes associated with massive particles for all junction orientations except when the junction is perpendicular to the deformation direction. 

Finally, transitions from KT (SKT) to AKT (ASKT) can be induced either by rotating the junction in the semi-Dirac phase or by applying a continuous uniaxial deformation parallel to it.

Our results provide a more comprehensive understanding of tunneling properties in materials with an anisotropic electronic
structure paving the way for advanced applications in electronic nanodevices and electron optics.

We expect that our predictions can be experimentally tested using a two-dimensional phononic crystal with a triangular lattice, as proposed in Ref. \cite{Zhu2023}. By introducing controlled anisotropy, as in the microwave graphene analogue \cite{Bellec}, a deformed dice lattice ($\alpha=1$) can be realized.

\section*{Acknowledgments}
We gratefully acknowledge J.-N. Fuchs for helpful discussions and for a critical reading of the manuscript.

\appendix

\section{Transmission probability and conductance for a junction parallel to the deformation direction}
In this appendix, we calculate the total transmission probability and the conductance for a junction parallel to the deformation axis.
As discussed in the main text, the transmission probability depends on the topology of the Fermi surfaces in the two regions, as illustrated in Fig. \ref{figtrans}, which can be summarized in three situations:

(i) \textit{Disconnected Fermi surfaces} ($\Delta/E<-1$):
Applying the matching conditions (Eq. (\ref{match cond gen})) with $\beta=0$ at $x=0$ for the total wave function given from Eq. (\ref{psi4 np}), we obtain a system of four equations
\be
\begin{aligned}
&1+r_{s}^{s}+r_{-s}^{s}=-t_{s}^{s}-t_{-s}^{s},\\ 
&a_{s}+a_{-s}r_{s}^{s}+a_{s}r_{-s}^{s}=a_{-s}t_{s}^{s}+a_{s}t_{-s}^{s},\\ 
&k_{s}+k_{-s}r_{s}^{s}-k_{s}r_{-s}^{s}=-k_{-s}t_{s}^{s}+k_{s}t_{-s}^{s},\\ 
&a_{s}k_{s}+a_{-s}k_{-s}r_{s}^{s}-a_{s}k_{s}r_{-s}^{s}=a_{-s}k_{-s}t_{s}^{s}-a_{s}k_{s}t_{-s}^{s},
\end{aligned}
\label{system np}
\ee
where $k_s=k_{x}^{+s}$ and
\be
a_{s}=  \cos^2 \varphi e^{-i \theta_s}+\sin^2 \varphi e^{i \theta_s}.
\label{as}
\ee
The transverse wave vector $k_{x}^{ss'}$ and the angle $\theta_s$  are given from Eq. (\ref{kx}) and Eq.  (\ref{angles}), respectively. 
Solving these equations yields the transmission and reflection amplitudes.
\be
\begin{aligned}
&t_{s}^{s}=\frac{-k_{s}\left ( a_s-a_{-s} \right)}{a_{-s}\left ( k_{s}+k_{-s} \right )}, 
&t_{-s}^{s}=0,\\
&r_{s}^{s}=\frac{-k_{s}\left ( a_s+a_{-s} \right)}{a_{-s}\left ( k_{s}+k_{-s} \right )},
&r_{-s}^{s}=\frac{k_s-k_{-s}}{k_{s}+k_{-s}}.
\end{aligned}
\label{t np}
\ee
The transmission and reflection probabilities are obtained from the transverse probability current $J_\perp$ [Eq. (\ref{trans curr})] by setting $\beta = 0$, as

\be 
\begin{aligned}
&T_{-s}^{s}=0,\\
&T_{s}^{s}=\frac{\left | J_\perp\left[\psi_p^{s -s}(0^+)\right] \right |}{\left | J_\perp\left[\psi_n^{s s}(0^-)\right]\right |}=\left |\frac{k_{-s}}{k_{s }} \right |\left | t_{s}^{s} \right |^2, \\
&R_{s}^{s}= \frac{\left | J_\perp\left[\psi_n^{s -s}(0^-)\right] \right |}{\left | J_\perp\left[\psi_n^{s s}(0^-)\right]\right |}=\left |\frac{k_{-s}}{k_{s }}\right | \left | r_{s}^{s} \right |^2,\\
&R_{-s}^{s}= \frac{\left | J_\perp\left[\psi_n^{-s s}(0^-)\right] \right |}{\left | J_\perp\left[\psi_n^{s s}(0^-)\right]\right |}= \left | r_{-s}^{s} \right |^2.
\end{aligned}
\label{T np}
\ee
The total transmission probability in situation (i), $T_{(i)}=\sum_{s=\pm} T_s^s+T_{-s}^{s}$, is then given by
\be
T_{(i)}(E,\Delta,\varphi,k_\parallel)=\frac{8k_+k_-}{\left ( k_++k_- \right)^2}\frac{1}{1+\cos^2 2\varphi \tan^2 \theta_+}.
\label{T(i)}
\ee

(ii) \textit{Connected Fermi surfaces} ($0<\Delta/E<1$):
Applying the matching conditions [Eq. (\ref{match cond gen})] with $\beta=0$ at $x=0$ for the total wave function given in Eq. (\ref{psi2 np}), we obtain the following system of four equations:
\be
\begin{aligned}
&1+r+A=-t-B,\\ 
&a_{+}+a_{+}r+a_{-}A=a_{+}t+a_{-}B,\\ 
&k_{+}-k_{+}r-k_{-}A=-k_{+}t-k_{-}B,\\ 
&a_{+}k_{+}-a_{+}k_{+}r-a_{-}k_{-}A=-a_{+}k_{+}t+a_{-}k_{-}B.
\end{aligned}
\label{system npii}
\ee
Solving these equations, we obtain the transmission and reflection amplitudes
\be
\begin{aligned}
&t=\frac{-ik_{+}\kappa \left ( a_-^2-a_{+}^2 \right)}{a_{-}a_{+}\left ( k_{+}^2+\kappa^2 \right )}, 
&r=\frac{k_{+}^2-\kappa^2}{k_{+}^2+\kappa^2}+\frac{ik_{+}\kappa\left (a_+^2+a_{-}^2 \right)}{a_{-}a_+\left (k_{+}^2+\kappa^2\right )}.
\end{aligned}
\label{t npii}
\ee

As in situation (i), the transmission and reflection probabilities are given by

\be 
T=\frac{\left | J_\perp\left[\psi_p^{-+}(0^+)\right] \right |}{\left | J_\perp\left[\psi_n^{++}(0^-)\right]\right |}=\left | t\right |^2,
R=\frac{\left | J_\perp\left[\psi_n^{-+}(0^-)\right] \right |}{\left | J_\perp\left[\psi_n^{++}(0^-)\right]\right |}=\left | r\right |^2.
\label{Tnp (ii)}
\ee 
The total transmission probability in situation (ii), $T_{(ii)}=T$, reads
\be 
T_{(ii)}(E,\Delta,\varphi,k_\parallel)=\frac{16k_+^2\kappa^2}{\left ( k_+^2+\kappa^2\right)^2}
\frac{\cos^2 2\varphi\tan^2 \theta_+}{\left(1+\cos^2 2\varphi\tan^2 \theta_+\right)^2},
\label{T(ii)}
\ee
where the reel term $\kappa=-ik_-$.

(iii) \textit{Partially connected Fermi surfaces} ($-1 < \Delta/E < 0$):\\
The total transmission probability can be deduced from situations (i) and (ii) as
\be 
T_{(iii)}=\begin{cases}
T_{(i)} \hspace{2cm} k_{in}<\left | k_{\parallel} \right |<k_{max} \vspace{0.2cm}\\
T_{(ii)}  \hspace{2cm} \left | k_{\parallel} \right |<k_{in}
\end{cases},
\label{Txnp}
\ee
where $k_{max}=\frac{E}{\hbar v_y}$ and $k_{in}=\frac{\sqrt{E^2-\Delta^2}}{\hbar v_y}$.
Here, $k_{in}$ represents the transverse momentum at which the Fermi surfaces in the two regions transition from being disconnected to connected (see Fig. \ref{figtrans}). For $|k_\parallel| < k_{in}$, the Fermi surfaces are connected, and the transmission follows $T_{(ii)}$, whereas for $|k_\parallel| > k_{in}$, the Fermi surfaces are disconnected, and the transmission follows $T_{(i)}$.

The conductance, given in the main text by Eq. (\ref{conductancex dirac}), can be expressed as
\be
G_x=4G_o\frac{L_y}{2\pi} \int _{0}^{k_{m}} T(k_{\parallel})dk_{\parallel}, 
\label{conducx}
\ee
where $L_y$ is the width of the sample in the $y$ direction.
The transmission probability $T$ and the upper integration limit $k_{m}$ depend on the ratio $\Delta/E$, i. e. on situations (i)-(iii). Accordingly, the conductance in the three situations reads
\be 
G_x=4G_o\frac{L_y}{2\pi}\begin{cases}
 \int _{0}^{k_{max}} T_{(i)}(k_{\parallel})dk_{\parallel} \hspace{3.1cm} \text{(i)}\vspace{0.2cm}\\
 \int _{0}^{k_{in}} T_{(ii)}(k_{\parallel})dk_{\parallel} \hspace{3.25cm}\text{(ii)}\vspace{0.25cm}\\
 \int _{0}^{k_{in}} T_{(ii)}(k_{\parallel})dk_{\parallel}+\int _{k_{in}}^{k_{max}} T_{(i)}(k_{\parallel})dk_{\parallel} \hspace{0.2cm}\text{(iii)}\vspace{0.2cm}\\
\end{cases}
\label{condx}
\ee 
Finally, the conductance $G_x^{SKT}$ in the SKT regime in the Dirac phase is obtained from Eq. (\ref{condx}) in situation (i) by setting $T_{(i)}(k_\parallel) = 2$ for all $k_\parallel$, giving
\be
G_x^{SKT}=4G_o L_y k_{max}/\pi.
\label{condskt}
\ee
Here, $T_{(i)}(k_\parallel) = 2$ reflects perfect transmission in the SKT regime, which arises from pseudospin conservation in the Dirac phase, ensuring that all transverse modes contribute fully to the conductance.
\label{A}

\section{The matching condition}
Here, we derive the general matching condition at the interface between the  $n$ and $p$ regions. The $np$ junction is oriented at an angle $\beta$ with respect to the deformation direction ($y$ direction).
Starting from the eigenvalue equation $\left[H+V\left ( x_\perp,x_\parallel \right )\right]\Psi \left ( x_\perp,x_\parallel \right )=E \Psi \left ( x_\perp,x_\parallel \right )$, where $x_\perp$ ($x_\parallel$) denotes the direction perpendicular (parallel) to the $np$ junction and $H$ is the Hamiltonian given in Eq. (\ref{unv_hamlt}), with the momentum components transformed as  
\be
\begin{array}{cc}
k_x=\cos\beta k_{\perp}-\sin\beta k_{\parallel},\\
k_y=\sin\beta k_{\perp}+\cos\beta k_{\parallel}.
\end{array}
\label{kxky}
\ee
The potential step $V\left ( x_\perp,x_\parallel \right )$ is given by
\be 
 V(x_\perp,x_\parallel)=V_o \Theta(x_\perp),  
\label{pot xperp2}
\ee
where $\Theta(x_\perp)$ is the Heaviside step function. 
Since the potential is uniform along the $x_\parallel$ direction, the parallel component of the wave vector, $k_\parallel$, is conserved. Consequently, the wave function can be expressed as $\Psi \left ( x_\perp,x_\parallel \right )=\psi \left (x_\perp \right )e^{ik_{\parallel}x_\parallel}$.

In the first step, we evaluate the integral of the eigenvalue equation over the interval $[-\epsilon , \epsilon]$, replacing $k_\perp$ with $-i \partial_{x_\perp}$. Taking the limit $\epsilon \to 0$, we obtain
\begin{equation}
\left(K_{\beta,\varphi}+M_{\beta,\varphi}\partial_{x_\perp}\right) \psi \left ( 0^+ \right )=\left(K_{\beta,\varphi}+M_{\beta,\varphi}\partial_{x_\perp}\right) \psi \left ( 0^- \right ),
\label{match cond1}
\end{equation} 
where 
$K_{\beta,\varphi}=i\left(\frac{\hbar^2}{2m}\sin2\beta k_{\parallel}S_x^\varphi-\hbar v_y \sin\beta S_y^\varphi\right)$ and
$M_{\beta,\varphi}=-\frac{\hbar^2}{2m}\cos^2\beta S_x^\varphi$. 

In the second step, we evaluate the double integral of the eigenvalue equation over the same interval $[-\epsilon , \epsilon]$. Taking the limit $\epsilon \to 0$ leads to
\begin{equation}
S_x^\varphi \psi \left ( 0^+ \right )=S_x^\varphi \psi \left ( 0^- \right ),
\label{match cond2}
\end{equation} 
where the pseudospin matrices $S_{x,y}^\varphi$ are given by  Eq. (\ref{pseudospinxy}).
By combining these two results, we obtain the full set of matching conditions at the $np$ junction
\begin{equation}
\begin{array}{cccc}
\psi_B(0^+)=\psi_B(0^-),\\
\partial_{x_\perp}\psi_B(0^+)=\partial_{x_\perp}\psi_B(0^-),\\
F_{\varphi}^+(0^+)=F_{\varphi}^+(0^-),\\
G_{\varphi,\beta}(0^+)=G_{\varphi,\beta}(0^-),
\end{array}
\label{match cond gen}
\end{equation}
where 
\be
\begin{aligned}
&G_{\varphi,\beta}(x_\perp)=\frac{\hbar}{m}\cos^2\beta \partial_{x_\perp}F_\varphi^+(x_\perp)-2 v_y \sin\beta F_\varphi^-(x_\perp),\\ 
&F_\varphi^{\pm}(x_\perp)=\cos\varphi \psi_A(x_\perp)\pm \sin\varphi \psi_C(x_\perp),
\end{aligned}
\ee
and $\psi_A$, $\psi_B$ and $\psi_C$ are the components of the wave function.

In the special case $\beta = \pi/2$, the matching condition follows solely from Eq. (\ref{match cond1}) and takes the form $S_y^\varphi \psi \left ( 0^+ \right )=S_y^\varphi \psi \left ( 0^- \right )$. Explicitly, this condition reads
\be
\begin{array}{cccc}
\psi_B(0^+)=\psi_B(0^-),\\
F_{\varphi}^-(0^+)=F_{\varphi}^-(0^-).\\
\end{array}
\label{match_pi2} 
\ee

Next, we show that these matching conditions guarantee the conservation of the probability current $J_\perp$, perpendicular to the $np$ junction. 
Starting from the Schr\"odinger equation $H\Psi=i\hbar \partial_t \Psi$ and using the probability conservation equation $\partial_t \left | \Psi \right |^2+\mathbf{\nabla.J}=0 $, we obtain
\begin{equation}
\begin{split}
J_\perp\left[\psi(x_\perp)\right]=&-\frac{\hbar}{m}\cos^2\beta \mathrm{Im} \left(F_{\varphi}^+(x_\perp)\partial_{x_\perp} \psi_B^*(x_\perp)\right)\\&
-\frac{\hbar}{m}\sin2\beta k_\parallel\mathrm{Re} \left(F_{\varphi}^+(x_\perp)\psi_B^*(x_\perp)\right)\\&
+\mathrm{Im}\left(G_{\varphi,\beta}(x_\perp)\psi_B^*(x_\perp)\right).
\end{split}
\label{trans curr}
\end{equation}
From this result, we see that the matching conditions [Eqs. (\ref{match cond gen}) and (\ref{match_pi2})] ensure the conservation of the probability current across the interface, i. e. $J_\perp\left[\psi(0^+)\right]=J_\perp\left[\psi(0^-)\right]$, at the interface $x_\perp=0$.
\label{B}

\section{The wave functions in the semi-Dirac phase}
\label{C}
In this appendix, we determine the wave function in the semi-Dirac phase ($\Delta=0$) as a function of the angle $\beta$. For an energy $E=V_o/2$, a transverse momentum $k_{\parallel}$ and an orientation angle $\beta$, the longitudinal momentum $k_\perp$ admits four possible solutions obtained from
\be
E^2=|f_{\lambda=2}(k_{x},k_{y})|^2,
\label{eq_kperp}
\ee
where $f_{\lambda}(k_{x},k_{y})$ is defined in Eq. (\ref{eqfk univ}) and the momentum components $(k_x,k_y)$ are expressed in terms of $(k_\perp,k_\parallel)$ through Eq. (\ref{kxky}). 
More explicitly, for $\beta \neq 0,\pi/2$, Eq.(\ref{eq_kperp}) reduces to the quartic equation
\be
k_x^4-pk_x^2-qk_x-r=0,
\label{trans moment sd}
\ee
where $p=-\frac{\hbar^2v_y^2}{(\hbar^2/(2m))^2}\tan^2\beta$, $q=-\frac{2\hbar^2v_y^2 k_\parallel}{(\hbar^2/(2m))^2}\frac{\tan\beta}{\cos\beta}$ and $r=-\frac{\hbar^2v_y^2 k_\parallel^2}{(\hbar^2/(2m))^2\cos^2\beta}+\frac{E^2}{(\hbar^2/(2m))^2}$. Using the identity $k_x^4=(k_x^2+t)^2-2kx^2t-t^2$, Eq. (\ref{trans moment sd}) can be rewritten as $(k_x^2+t)^2=(2t+p)k_x^2+qk_x+r+t^2$. We now choose $t$ such that the discriminant of the right-hand side vanishes. This condition yields a cubic equation for $t$
\be
(t+p/6)^3+3v(t+p/6)+2u=0,
\ee
where $v=r/3-p^2/36$ and $u=p^3/216+rp/6-q^2/16$.
Applying Cardano’s method, one obtains
\be
t=\left[-u+\sqrt{u^2+v^3}\right]^{1/3}+\left[-u-\sqrt{u^2+v^3}\right]^{1/3}-p/6.
\ee
The four solutions of Eq.~(\ref{trans moment sd}) can then be expressed as 
\be
k_x^{ss'}=\frac{s}{2}\sqrt{2t+p}+\frac{s'}{2}\sqrt{p-2t+2sq/\sqrt{p+2t}},
\ee    
\label{kx_ssp}
where $s,s'=\pm 1$ label the four wave vectors.\\
Finally, the transverse wave vectors $k_\perp^{ss'}$, solution of Eq. (\ref{eq_kperp}), are given by
\be 
k_\perp^{ss'}=\frac{k_x^{ss'}}{\cos\beta}+k_\parallel \tan\beta.
\label{kperp_ssp}
\ee
The corresponding normalized eigenstate is of the form
\be 
\psi_{\nu}^{(s,s')}(x_\perp)= \frac{1}{\sqrt{2}} \begin{pmatrix}
\cos\varphi e^{-i\theta^{ss'}} \\ 
\nu\\ 
\sin\varphi e^{i\theta^{ss'}} 
\end{pmatrix}e^{ik_{\perp}^{ss'}x_\perp},
\label{psi sd}
\ee 
where $\nu=\pm 1$ is the band index and $\theta^{ss'}$ is the phase defined through
\be 
e^{i\theta^{ss'}}=\frac{f_{\lambda=2}(k_{x}^{ss'},k_{y}^{ss'})}{|f_{\lambda=2}(k_{x}^{ss'},k_{y}^{ss'})|},
\label{angles sd}
\ee
where $k_y^{ss'}=\frac{k_\parallel +\sin\beta k_x^{ss'}}{\cos\beta}$.
It can be verified that $k_\perp^{ss'}$ is real when $s=-\mathrm{sign}(k_\parallel)$, corresponding to a propagating mode, and imaginary when $s=\mathrm{sign}(k_\parallel)$, corresponding to an evanescent mode.

For $\beta=0$ the wave function is obtained from Eq. (\ref{psix np}) with $\Delta=0$.
In the case $\beta=\pi/2$, Eq.~(\ref{eq_kperp}) admits two real solutions,
\be
k_{\perp}^{s}= \frac{s}{\hbar v_y} \sqrt{E^2-\left(\frac{\hbar^2k_{\parallel}^2}{2m}\right)^2}, 
\label{kqy}
\ee
where $s=\pm 1$ denotes the two possible longitudinal momenta.
These solutions correspond to propagating waves provided that $E^2-\left(\frac{\hbar^2k_{\parallel}^2}{2m}\right)^2>0$.
The associated normalized eigenstate reads
\be 
\psi_{\nu}^{s}(x_\perp)= \frac{1}{\sqrt{2}} \begin{pmatrix}
\cos\varphi e^{-i\theta_s} \\ 
\nu\\ 
\sin\varphi e^{i\theta_s} 
\end{pmatrix}e^{ik_{\perp}^{s}x_\perp},
\label{fun y}
\ee 
where $\nu=\pm$ denotes the band index, and the phase angle is given by
\be 
\theta_{s}=\arg \left [\frac{\hbar^2k_{\parallel}^2}{2m}+i s \sqrt{E^2-\left(\frac{\hbar^2k_{\parallel}^2}{2m}\right)^2}\right ].
\label{anglesy}
\ee

\section{Tunneling properties in the Dirac phase}
 \label{DP}
 \renewcommand{\thefigure}{D-\arabic{figure}}
 \setcounter{figure}{0}
 In this appendix, we investigate the tunneling across the $\mathit{np}$ junction as a function of its orientation $\beta$ relative to the deformation axis, in the Dirac phase ($0<E<<-\Delta$).
 
In the vicinity of the Dirac point $D_{+}$, the Hamiltonian can be expressed as (see Eqs (\ref{eq1}) and (\ref{eqfk dirac}))
\be
H=\hbar \left(v_{\perp}S_{\perp}^\varphi \delta k_{\perp}+ v_{\parallel}S_{\parallel}^\varphi \delta k_{\parallel} \right),
\label{hamiltonian dirac b}
\ee
where 
\be
\begin{aligned}
&\delta k_{\perp}=\cos\beta \delta k_{x}+\sin\beta \delta k_{y},\\
&\delta k_{\parallel}=-\sin\beta \delta k_{x}+\cos\beta \delta k_{y}.
\end{aligned}
\label{convert deltak}
\ee
The pseudospin matrices read
 \be
 S_{\chi=\perp, \parallel}^{\varphi}=\begin{pmatrix}
0 &\cos\varphi e^{-i\gamma_{\chi}}  &0 \\ 
\cos\varphi e^{i\gamma_{\chi}}& 0 &\sin\varphi e^{-i\gamma_{\chi}} \\ 
 0& \sin\varphi e^{i\gamma_{\chi}}&0 
\end{pmatrix},
\label{pseudospin} 
\ee
 where $v_{\perp}e^{i\gamma_{\perp}}=v_x \cos \beta+i v_y \sin \beta$ and $v_{\parallel}e^{i\gamma_{\parallel}}=-v_x \sin \beta+i v_y \cos \beta$.
The low-energy spectrum consists of an anisotropic Dirac cone with energy $E = \nu \hbar \sqrt{v_x^2 \delta k_x^2 + v_y^2 \delta k_y^2}$ and a flat band at energy $E = 0$, where $\nu = \pm$ denotes the band index.

For a fixed energy $E = V_o/2$ and transverse momentum $\delta k_{\parallel}$, we obtain two possible longitudinal momenta $\delta k_{\perp}^{s}$ given by
\be
\begin{split}
\delta k_{\perp}^{s}=&-\frac{v_{\parallel}}{v_{\perp}}  \cos(\gamma_{\parallel}-\gamma_{\perp})\delta k_{\parallel}\\&+ \frac{s}{\hbar v_{\perp}} \sqrt{E^2 -\left(\hbar v_{\parallel}\sin(\gamma_{\parallel}-\gamma_{\perp})\delta k_{\parallel}\right)^2},
\label{dkperp}
\end{split}
\ee
where $s = \pm 1$ labels the two solutions for the longitudinal momentum.
The corresponding normalized eigenstate perpendicular to the $\mathit{np}$ junction (along the $x_\perp$-direction) is given by
\be 
\Psi_{\nu}^{s}(x_\perp)= \frac{1}{\sqrt{2}} \begin{pmatrix}
\cos\varphi e^{-i\theta_{s}} \\ 
\nu\\ 
\sin\varphi e^{i\theta_{s}} 
\end{pmatrix}e^{i\delta k_{\perp}^{s}x_\perp},
\label{psi dirac}
\ee 
where 
\be 
\theta_{s}=\arg \left(v_{\perp}e^{i\gamma_{\perp}}\delta k_{\perp}^s+v_{\parallel}e^{i\gamma_{\parallel}}\delta k_{\parallel}\right)
\label{angles dirac}
\ee
is the pseudospin angle. 
The propagation direction is determined from the group velocity, whose components are given by
\be
\begin{aligned}
&V_{\perp \nu}^s=\frac{1}{\hbar}\frac{\partial E }{\partial \delta k_{\perp}^s}=\nu v_{\perp}\cos \left( \theta_{s}-\gamma_{\perp}\right),\\
&V_{\parallel \nu}^{s}=\frac{1}{\hbar}\frac{\partial E }{\partial\delta k_{\parallel}}=\nu v_{\parallel}\cos \left( \theta_{s}-\gamma_{\parallel}\right).
\end{aligned}
\label{velocity dirac}
\ee
The total wave function in both spatial regions takes the form
\be
\begin{aligned}
&\Psi(x_\perp<0)=\Psi_{i}(x_\perp)+r \Psi_{r}(x_\perp), \\ 
&\Psi(x_\perp>0)= t \Psi_{t}(x_\perp),  
\end{aligned}
\label{trans dirac}
\ee 
where $\Psi_{i}$, $\Psi_{r}$ and $\Psi_{t}$ are the incident, reflected and transmitted wave functions, respectively. 
To determine the matching conditions, we integrate the eigenvalue equation  $\left [H+V(x_\perp)\right ]\Psi (\xi) =E \Psi(x_\perp)$ over the interval $[-\varepsilon ,\varepsilon]$, substituting $\delta k_{\perp}=-i\partial_{x_\perp}$, and then taking the limit $\varepsilon \rightarrow 0$. This yields the matching condition
\be 
S_{\perp}^\varphi \Psi(0^+)= S_{\perp}^\varphi \Psi(0^-),
\label{cond dirac} 
\ee
which can be written as
\be
\begin{array}{cc}
\psi_B(0^-)=\psi_B(0^+)\\
c_1\psi_A(0^-)+c_2\psi_C(0^-)=c_1\psi_A(0^+)+c_2\psi_C(0^+)
\end{array}
 \ee
where $c_1=\cos \varphi (v_x \cos \beta+ iv_y \sin \beta)$ and $c_2=\sin \varphi (v_x \cos \beta- iv_y \sin \beta)$.
For the particular case of the dice lattice ($\alpha=1$), these boundary conditions reduce to those derived in Ref. \cite{Iurov2020}, upon choosing the corresponding velocity parameters $v_x$ and $v_y$.
 Applying this matching condition, we obtain the reflection and the transmission amplitudes
\be
r=-\frac{u_i+u_t}{u_r+u_t}, \hspace{1cm} t=\frac{u_i-u_r}{u_r+u_t},
\label{r t dirac}
\ee
where 
\be
u_\chi=\cos^2\varphi e^{-i\left(\theta_\chi-\gamma_{\perp}\right)}+\sin^2\varphi e^{i\left(\theta_\chi-\gamma_{\perp}\right)}, 
\label{u_chi}
\ee
To evaluate the transmission probability, we first introduce the transverse component of the probability current, $J_\perp$.
Starting from the Schr\"odinger equation $H\Psi=i\hbar \partial_t \Psi$ and using the probability conservation equation $\partial_t \left | \Psi \right |^2+\mathbf{\nabla.J}=0 $, we find
\be
J_{\perp}[\Psi]=v_{\perp} \Psi^\dagger S_{\perp}^\varphi \Psi, 
\label{current dirac}
\ee
where $\Psi=\left ( \psi_A,\psi_B,\psi_C \right )^t$. 
Note that the matching condition [Eq. (\ref{cond dirac})] corresponds to the conservation of the probability current perpendicular to the $np$ junction.
The transmission probability as a function of the incidence angle $\phi_i$ is given by 
\be 
T=\frac{\left | J_\perp\left[\Psi_t\right] \right |}{\left | J_{\perp}\left[\Psi_i\right]\right |}|t|^2=\frac{1}{1+\frac{v_{\perp}^4}{v_x^2v_y^2}\cos^22\varphi \left(\tan \phi_i-\tan \phi_o\right)^2},
\label{T dirac}
\ee 
where $\tan \phi_o=\frac{v_{\parallel}}{v_{\perp}}  \cos(\gamma_{\parallel}-\gamma_{\perp})$. 
Note that the above transmission probability coincides with the result reported in Ref.~\cite{Betancurprb2018} for graphene ($\alpha = 0$).
When $\phi_i = \phi_o$, the KT occurs for any values of $\alpha$ and $\beta$: perfect transmission is achieved at normal incidence only for $\beta = 0$ and $\beta = \pi/2$, and at oblique incidence for other $\beta$ values.
A notable special case is the dice lattice ($\alpha = 1$), where perfect transmission holds for all incidence angles, giving rise to the SKT effect for any junction orientation $\beta$.
 In all cases, the perfect transmission arises from pseudospin conservation, as explained below.
Following the approach of Ref. \cite{Zhang2018}, the perfect transmission occurs when $r=0$ which amounts to the conditions $u_i=-u_{t} \neq u_r$. Using Eq. (\ref{u_chi}), we obtain the conditions of perfect transmission 
\be
\begin{aligned}
&\cos(\theta_i-\gamma_\perp)+\cos(\theta_t-\gamma_\perp)=0, \\ 
&\cos 2\varphi \left(\sin(\theta_i-\gamma_\perp)+\sin(\theta_t-\gamma_\perp)\right)=0. 
\end{aligned}
\ee 
The first condition expresses the conservation of the transverse component of the pseudospin, $\left \langle S_{\perp} \right \rangle_{i}=\left \langle S_{\perp} \right \rangle_{t}$, where $\left \langle S_{\perp} \right \rangle_{\chi}=\expval{S_{\perp}^{\varphi}}{\Psi_{\chi}}$.
In the case where $\phi_i=\phi_o$, we have $\sin(\theta_i-\gamma_\perp)=\sin(\theta_t-\gamma_\perp)=0$, which corresponds to pseudospin conservation between the incident and transmitted waves in the KT regime. In the dice lattice, however, the SKT effect is instead associated with the conservation of the transverse pseudospin component.

\begin{figure}[h!]
  \centering
\setlength{\unitlength}{1mm}
\includegraphics[width=0.48\textwidth]{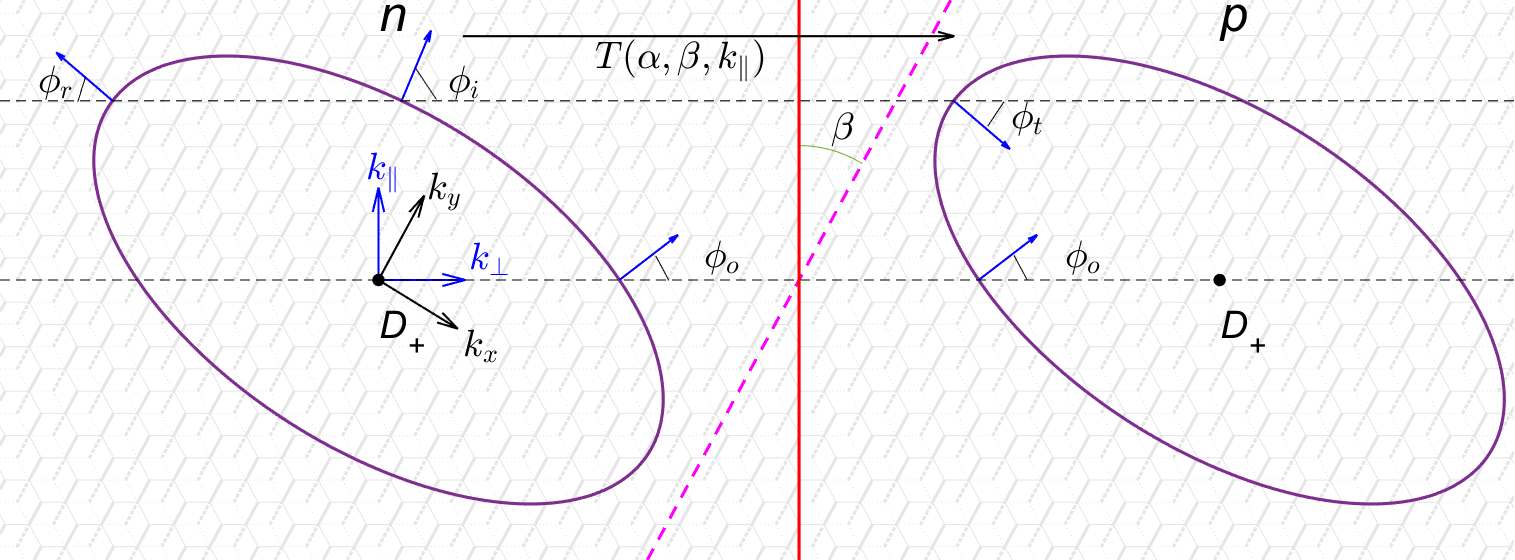}
\includegraphics[width=0.5\textwidth]{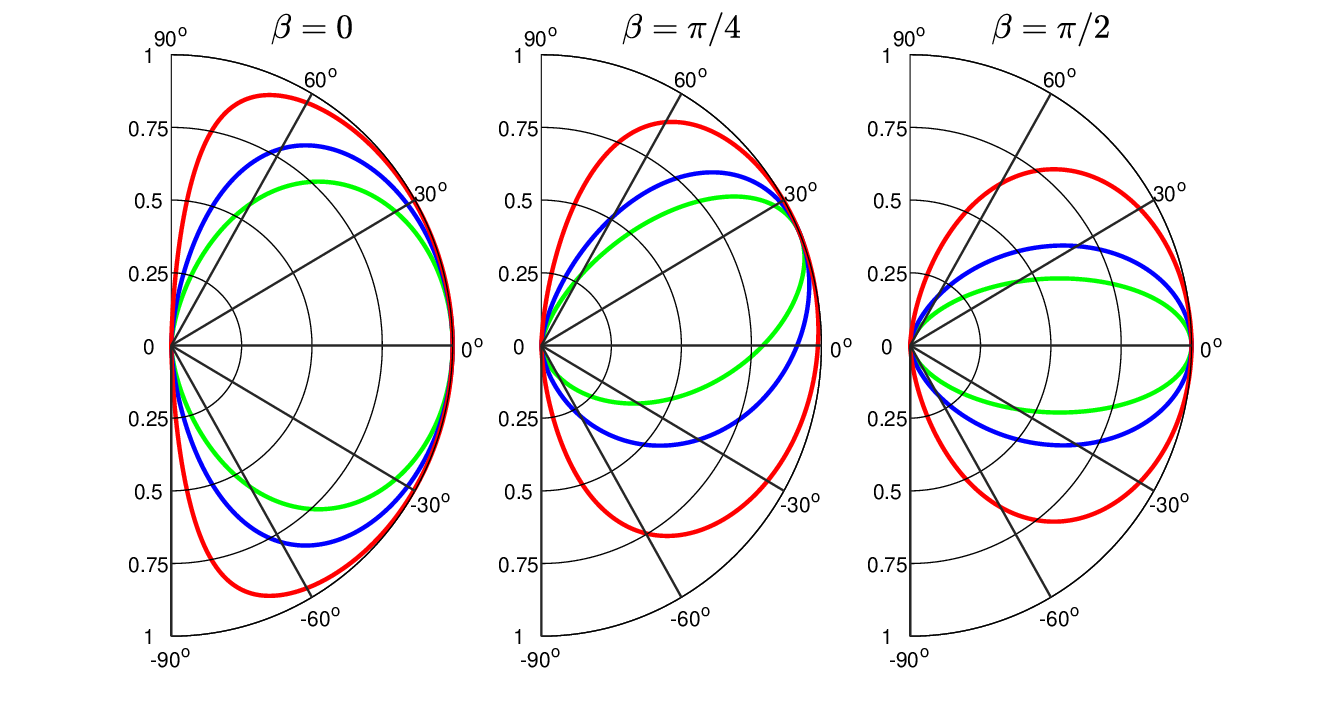}

\caption{(Color online) 
Upper panel: Schematic illustration of the transmission probability in the Dirac phase across a $pn$ junction oriented with an angle $\beta=\pi/6$ with respect to the deformation axis (magenta dashed line). The blue arrows indicate the group velocities. 
The horizontal dashed lines correspond to the conserved wave vector component $\delta k_{\parallel}$.
When $\delta k_{\parallel}=0$ (lower dashed line), KT occurs at an oblique incidence angle $\phi_o$. 
Lower panel: Polar plot of the transmission probabilities as a function of the incidence angle for various values of $\beta$  and $\alpha$ (green for $\alpha=0$, blue for $\alpha=0.5$ and red for $\alpha=0.8$). The deformation strength is set to $\lambda=1.5$.}
\label{plotdirac}
\end{figure}

For a fixed $\alpha$, the junction transparency increases with $\beta$, and for larger $\alpha$ it becomes even more transparent (see Fig. \ref{plotdirac}), consistent with the behavior of the undeformed $\alpha-T_3$ lattice \cite{{Urban},{Illes2017}}.

We conclude that the electronic transport properties in the Dirac phase of the deformed $\alpha-T_3$ lattice are similar to those of the undeformed case. This similarity arises from the fact that particles behave as massless Dirac fermions in this phase.  

\section{Tunneling properties along the deformation direction}
 \label{KT y}

In this Appendix, we investigate the tunneling properties when the $np$ junction is perpendicular to the deformation direction. 
In this case, the potential is given by
\be 
 V(y)=V_o \Theta(y).  
\label{pot x}
\ee
where $\Theta(y)$ is the Heaviside step function.\\
The potential step is uniform along the $x$ direction so that the $x$ component of the wave vector is conserved.
The wave function can thus be written as $\Psi \left ( x,y \right )=\psi \left ( y \right )e^{ik_x x}$. 
For a given energy $E=V_o/2$, transverse momentum $k_{x}$ and a gap parameter $\Delta$, there exist two longitudinal momenta $k_{y}^{s}$ that satisfy the relation $E^2 = |f_{\lambda}(k_x, k_y)|^2$. They are given by 
\be
k_{y}^{s}= \frac{s}{\hbar v_y} \sqrt{E^2-\left(\Delta +\frac{\hbar^2k_{x}^2}{2m}\right)^2},
\label{kqy}
\ee
where $s=\pm$ denotes the two longitudinal momenta, and $f_{\lambda}(k_{x},k_{y})$ is given in Eq. (\ref{eqfk univ}). 
This wave vector is real, corresponding to a propagating mode, when $E^2-\left(\Delta +\frac{\hbar^2k_{x}^2}{2m}\right)^2>0$.
The corresponding normalized eigenstate is 
\be 
\psi_{\nu}^{s}(y)= \frac{1}{\sqrt{2}} \begin{pmatrix}
\cos\varphi e^{-i\theta_s} \\ 
\nu\\ 
\sin\varphi e^{i\theta_s} 
\end{pmatrix}e^{ik_{y}^{s}y}
\ee 
\label{fun y}
where $\nu=\pm$ is the band index and 
\be 
\theta_{s}=\arg \left [\left(\Delta +\frac{\hbar^2k_{x}^2}{2m}\right)+i \hbar v_y k_y^s \right],
\label{anglesy}
\ee
which determines the in-plane pseudospin orientation with respect to the $k_x-$axis.
The total wave function in the two regions of space is given by

\be
\begin{aligned}
&\psi(y<0)=\psi_{n}^{+}(y)+r \psi_{n}^{-}(y) \\ 
&\psi(y>0)= t \psi_{p}^{-}(y)  
\end{aligned}
\label{tr y np}
\ee 
where $r$ and $t$ are respectively the reflection and the transmission amplitudes. 
Applying the matching condition given by Eq. (\ref{match_pi2}), we obtain the reflection and transmission amplitudes
\be
r=-\frac{a_++a_-}{2a_-}, \hspace{1cm}     t=\frac{a_+-a_-}{2a_-},
\label{t npy}
\ee
where $a_{\pm}=\cos^2 \varphi e^{-i\theta_\pm}-\sin^2 \varphi e^{i\theta_\pm}$.
The transmission probability is given by $T=\frac{\left | J_\perp\left[\psi_p^-\right] \right |}{\left | J_{\perp}\left[\psi_n^+\right]\right |}|t|^2$ which reads
\be 
T=\frac{E^2-\left(\Delta +\frac{\hbar^2k_x^2}{2m}\right)^2}{E^2-\left(\Delta +\frac{\hbar^2k_x^2}{2m}\right)^2+\cos^22\varphi \left(\Delta +\frac{\hbar^2k_x^2}{2m}\right)^2}.
\label{T dirac y}
\ee 

Perfect transmission occurs at $k_x=\pm \sqrt{\frac{-2m\Delta}{\hbar^2}}$ with $\Delta \leq 0$, as a consequence of KT effect arising from pseudospin conservation. A similar perfect transmission is also observed in the dice lattice ($\alpha=1$), where it originates from SKT effect, associated with the conservation of the transverse pseudospin component.
Importantly, since particles remain massless Dirac fermions for all $\Delta \leq 0$, so both KT and SKT are unaffected by the deformation.


\begin{thebibliography}{99}
\bibitem{Klein1929} O. Klein, Die Reflexion von Elektronen an einem Potentialsprung
nach der relativistischen Dynamik von Dirac, Z. Phys. \textbf{53}, 157 (1929).
\bibitem{Calogeracos} A. Calogeracos, and N. Dombey, History and physics of the Klein paradox, Contemp. Phys. \textbf{40}, 313 (1999).
\bibitem{Dombey} N. Dombey, and A. Calogeracos, Seventy years of the Klein paradox, Phys. Rep. \textbf{315}, 41 (1999).
\bibitem{Novoselov} K. S. Novoselov, A. K. Geim, S. V. Morozov, D. Jiang, Y. Zhang, S. V. Dubonos, I. V. Grigorieva,
and A. A. Firsov, Electric Field Effect in Atomically Thin Carbon Films, Science \textbf{306}, 666 (2004).
\bibitem{Huard} B. Huard, J. A. Sulpizio, N. Stander, K. Todd, B. Yang, and D. Goldhaber-Gordon, Transport Measurements Across a Tunable Potential Barrier in Graphene, Phys. Rev. Lett. \textbf{98}, 236803 (2007).
\bibitem{Stander} N. Stander, B. Huard, and D. Goldhaber-Gordon, Evidence for Klein Tunneling in Graphene p-n Junctions, Phys. Rev. Lett. \textbf{102}, 026807 (2009).
\bibitem{Young2009} A.F. Young, and P. Kim, Quantum interference and Klein tunnelling in graphene heterojunctions, Nature Phys. \textbf{5}, 222 (2009).
\bibitem{Katsnelson} M. I. Katsnelson, K. S. Novoselov, and A. K. Geim, Chiral tunnelling and the Klein paradox
in graphene, Nature Phys. \textbf{2}, 620 (2006).
\bibitem{Allain} P.E. Allain, and J.N. Fuchs, Klein tunneling in graphene: optics with massless electrons, Eur. Phys. J. B\textbf{83}, 301 (2011).
\bibitem{Zhihao} Z. Lan, N. Goldman, A. Bermudez, W. Lu, and P. \"Ohberg, Dirac-Weyl fermions with arbitrary spin in two-dimensional optical superlattices, Phys. Rev. B \textbf{84}, 165115 (2011).
\bibitem{Urban} D.F. Urban, D. Bercioux, M. Wimmer, and W. H\"ausler, Barrier transmission of Dirac-like pseudospin-one particles, Phys. Rev. B \textbf{84}, 115136 (2011).
\bibitem{Betancur} Y. Betancur-Ocampo, G. Cordourier-Maruri, V. Gupta, and R. de Coss, Super-Klein tunneling of massive pseudospin-one particles, Phys. Rev. B \textbf{96}, 024304 (2017).
\bibitem{Fang} A. Fang, Z.Q. Zhang, S.G. Louie, and C.T. Chan, Klein tunneling and supercollimation of pseudospin-1 electromagnetic waves, Phys. Rev. B \textbf{93}, 035422 (2016).
\bibitem{Shen} R. Shen,L. B. Shao, B. Wang, and  D. Y. Xing, Single Dirac cone with a flat band touching on line-centered-square optical lattices, Phys. Rev. B \textbf{81}, 041410(R) (2010).
\bibitem{Wu2024} Hao Wu, Hailong He, Ze Dong, Liping Y, Weiyin Deng, Manzhu Ke, and Zhengyou Liu, Super Klein tunneling in phononic Lieb lattices, Phys. Rev. Applied \textbf{21}, 034026 (2024).
\bibitem{Kim2019} K. Kim, Super-Klein tunneling of Klein-Gordon particles, Results Phys. \textbf{12}, 1391 (2019).
\bibitem{Betancur2018} Y. Betancur-Ocampo, Controlling electron flow in anisotropic Dirac
materials heterojunctions: a super-diverging lens, J. Phys.: Condens. Matter \textbf{30}, 435302 (2018).
\bibitem{Sutherland} B. Sutherland, Localization of electronic wave functions due to local topology, Phys. Rev. B \textbf{34}, 5208 (1986).
\bibitem{Vidal} J. Vidal, R. Mosseri, and B. Dou¸cot, Aharonov-Bohm Cages in Two-Dimensional Structures, Phys. Rev. Lett. \textbf{81}, 5888 (1998).
\bibitem{Vidal2001} J. Vidal, P. Butaud, B. Douçot, and R. Mosseri, Disorder and interactions in Aharonov-Bohm cages, Phys. Rev. B \textbf{64}, 155306 (2001).
\bibitem{Raoux} A. Raoux, M. Morigi, J.-N. Fuchs, F. Pi\'{e}chon, and G. Montambaux, From Dia- to Paramagnetic Orbital Susceptibility of Massless Fermions, Phys. Rev. Lett. \textbf{112}, 026402 (2014).
\bibitem{Illes2016} E. Illes, and E. J. Nicol, Magnetic properties of the $\alpha-T_3$  model: Magneto-optical conductivity and the Hofstadter butterfly, Phys. Rev. B \textbf{94}, 125435 (2016).
\bibitem{Malcolm} J. D. Malcolm and E. J. Nicol, Magneto-optics of massless Kane fermions: Role of the flat band and unusual Berry phase, Phys. Rev. B \textbf{92}, 035118 (2015).
\bibitem{Illes2017} E. Illes, and E. J. Nicol, Klein tunneling in the $\alpha-T_3$ model, Phys. Rev. B \textbf{95}, 235432 (2017).
\bibitem{Cunha2022} S. M. Cunha, D. R. da Costa, J. M. Pereira, R. N. C. Filho, B. V. Duppen, and F. M. Peeters, Tunneling properties in $\alpha-T_3$ lattices: Effects of symmetry-breaking terms, Phys. Rev. B \textbf{105}, 165402 (2022).
\bibitem{Iurov2022} A. Iurov, L. Zhemchuzhna, G. Gumbs, D. Huang, and P. Fekete, Optically modulated tunneling current of dressed electrons in graphene and a dice lattice, Phys. Rev. B \textbf{105}, 115309 (2022).
\bibitem{Bouhadida2020} F. Bouhadida, L. Mandhour, and S. Charfi-Kaddour, Magnetic Fabry-Pérot interferometer for valley filtering in a honeycomb-dice model, Phys. Rev. B \textbf{102}, 075443 (2020).
\bibitem{Varlet2016} A. Varlet, M. H. Liu, D. Bischoff, P. Simonet, T. Taniguchi, K. Watanabe, K. Richter, T. Ihn, and K. Ensslin, Band gap and broken chirality in single-layer and bilayer graphene, Phys. Status Solidi \textbf{10}, 46 (2016).
\bibitem{Du} R. Du, M.H. Liu, J. Mohrmann, F. Wu, R. Krupke, H.V. L\"ohneysen, K. Richter, and R. Danneau, Tuning Anti-Klein to Klein Tunneling in Bilayer Graphene, Phys. Rev. Lett. \textbf{121}, 127706 (2017).
\bibitem{Bahat} O. Bahat-Treidel, O. Peleg, M. Grobman, N. Shapira, M. Segev, and T. Pereg-Barnea, Klein Tunneling in Deformed Honeycomb Lattices, Phys. Rev. Lett. \textbf{104}, 063901 (2010).
\bibitem{Liu2016} M.-H. Liu, J. Bundesmann, and K. Richter, Spin-dependent Klein tunneling in graphene: Role of Rashba spin-orbit coupling, Phys. Rev. B \textbf{85}, 085406 (2012).
\bibitem{Banerjee} S. Banerjee, and W. E. Pickett, Phenomenology of a semi-Dirac semi-Weyl semimetal, Phys. Rev. B \textbf{86}, 075124 (2012).
\bibitem{Yonatan} Y. Betancur-Ocampo, F. Leyvraz, and T. Stegmann, Electron Optics in Phosphorene pn Junctions: Negative Reflection
and Anti-Super-Klein Tunneling, Nano Lett. \textbf{19}, (11), 7760-7769 (2019).
\bibitem{Gilles1} G. Montambaux, F. Pi\'{e}chon, J.-N. Fuchs, and M. O. Goerbig, A universal Hamiltonian for the motion and the merging of Dirac cones
in a two-dimensional crystal, Eur. Phys. J. B \textbf{72}, 509 (2009).
\bibitem{Pereira} V. M. Pereira, A. H. Castro Neto, and N. M. R. Peres, Tight-binding approach to uniaxial strain in graphene, Phys. Rev. B \textbf{80}, 045401 (2009).
\bibitem{Kim2009} K. S. Kim, Y. Zhao, H. Jang, S. Y. Lee, J. M. Kim, K. S. Kim, J.-H. Ahn, P. Kim, J.-Y. Choi, and B. H. Hong, Large-scale pattern growth of graphene films for stretchable transparent electrodes, Nature \textbf{457}, 706 (2009).
\bibitem{Gail2009} R. de Gail, J. N. Fuchs, M. O. Goerbig, F. Piechon, and G. Montambaux, Manipulation of Dirac points in graphene-like crystals, Physica B \textbf{407}, 1948 (2012).
\bibitem{Gilles2} G. Montambaux, F. Pi\'{e}chon, J.-N. Fuchs, and M. O. Goerbig, Merging of Dirac points in a two-dimensional crystal, Phys. Rev. B \textbf{80}, 153412 (2009).
\bibitem{Tarruel} L. Tarruell, D. Greif, T. Uehlinger, G. Jotzu, and T. Esslinger, Creating, moving and merging Dirac points with a
Fermi gas in a tunable honeycomb lattice, Nature, \textbf{483}, 302 (2012). 
\bibitem{Feilhauer} J. Feilhauer, W. Apel, and L. Schweitzer, Merging of the Dirac points in electronic artificial graphene, Phys. Rev. B \textbf{92}, 245424 (2009).
\bibitem{Polini2013} M. Polini, F. Guinea, M. Lewenstein, H. C. Manoharan, and V. Pellegrini, Artificial honeycomb lattices for electrons,
atoms and photons, Nature Nanotech. \textbf{8}, 625 (2013).
\bibitem{Bellec} M. Bellec, U. Kuhl, G. Montambaux, and F. Mortessagne, Topological Transition of Dirac Points in a Microwave Experiment, Phys. Rev. Lett. \textbf{110}, 033902 (2013).
\bibitem{Montambaux2018} G. Montambaux, Artificial graphenes: Dirac matter beyond condensed matter, Comptes Rendus Phys. \textbf{19}, 285–305 (2018).
\bibitem{Shao2024} Y. Shao, S. Moon, A. N. Rudenko, J. Wang, J. Herzog-arbeitman, M. Ozerov, D. Graf, Z. Sun, R. Queiroz, S. H. Lee, Y. Zhu, Z. Mao, M. I. Katsnelson, B. A. Bernevig, D. Smirnov, A. J. Millis, and D. N. Basov, Semi-Dirac Fermions in a Topological Metal, Phys. Rev. X \textbf{14}, 41057 (2024).
\bibitem{Gutierrez2016} C. Guti\'errez, C.-J. Kim, L. Brown, T. Schiros, D. Nordlund,E. B. Lochocki, K. M. Shen, J. Park, and A. N. Pasupathy, Imaging chiral symmetry breaking from Kekulé bond order in graphene Nat. Phys. \textbf{12}, 950 (2016).
\bibitem{Iurov2023} A. Iurov, L. Zhemchuzhna, G. Gumbs, and D. Huang, Application of the WKB theory to investigate electron tunneling in Kek-Y graphene. Appl. Sci., \textbf{13}, 6095 (2023)
\bibitem{Mojarro2020} M. A. Mojarro, V. G. Ibarra-Sierra, J. C. Sandoval-Santana, R. Carrillo-Bastos, and G. G. Naumis, Dynamical floquet spectrum of kekulé-distorted graphene under normal incidence of electromagnetic radiation, Phys. Rev. B \textbf{102}, 165301 (2020).
\bibitem{Betancurprb2018} Y. Betancur-Ocampo, Partial positive refraction in asymmetric Veselago lenses of uniaxially strained graphene, Phys. Rev. B \textbf{98}, 205421 (2018).
\bibitem{Saha} K. Saha, R. Nandkishore, and S. A. Parameswaran, Valley-selective Landau-Zener oscillations in semi-Dirac p-n junctions, Phys. Rev. B \textbf{96}, 045424 (2017).
\bibitem{Adroguer} P. Adroguer, D. Carpentier, G. Montambaux, and E. Orignac, Diffusion of Dirac fermions across a topological merging transition in two dimensions, Phys. Rev. B \textbf{93}, 125113 (2016).
\bibitem{Shengyuan} Yee Sin Ang, Shengyuan A. Yang, C. Zhang, Zhongshui Ma, and L. K. Ang, Valleytronics in merging Dirac cones: All-electric-controlled valley filter, valve, and universal
reversible logic gate, Phys. Rev. B \textbf{96}, 245410 (2017).
\bibitem{Zhu2020} Hao-Fu Zhu, Xue-Qian Yang, Jun Xu,  and Shuai Cao, Barrier tunneling of quasiparticles in double-Weyl semimetals, Eur. Phys. J. B \textbf{93}, 4 (2020).
\bibitem{Donck} M. Van der Donck, F. M. Peeters, and B. Van Duppen, Transport properties of bilayer graphene in a strong in-plane magnetic field, Phys. Rev. B \textbf{93}, 115423 (2016).
\bibitem{Anna} L Dell'Anna, P Majari, and M R Setare, From Klein to anti-Klein tunneling in graphene tuning the Rashba spin-orbit
interaction or the bilayer coupling, J. Phys.: Condens. Matter \textbf{30} 415301 (2018).
\bibitem{Piechon} F. Pi\'{e}chon, J.-N. Fuchs, A. Raoux, and G. Montambaux, Tunable orbital susceptibility in $\alpha-T_3$ tight-binding models, J. Phys.: Conf. Ser. \textbf{603}, 012001 (2015).
\bibitem{Iurov2019} A. Iurov, G. Gumbs, and D. Huang, Peculiar electronic states, symmetries, and Berry phases in irradiated $\alpha-T_3$ materials, Phys. Rev. B \textbf{99}, 205135 (2019). 
\bibitem{Lifshitz} I.M. Lifshitz, Anomalies of electron characteristics of a metal in the high pressure region, Sov. Phys. JETP \textbf{11}, 1130 (1960).
\bibitem{Dey2019} B. Dey, and T. Kanti Ghosh, Floquet topological phase transition in the $\alpha-T_3$ lattice, Phys. Rev. B \textbf{99}, 205429 (2019).
\bibitem{Blanter} Ya.M. Blanter, and M. B\"uttiker, Shot noise in mesoscopic conductors, Phys. Rep. \textbf{336}, 1 (2000).
\bibitem{Dey2018} B. Dey and T. Kanti Ghosh, Photoinduced valley and electron-hole symmetry breaking in $\alpha-T_3$ lattice:
The role of a variable Berry phase, Phys. Rev. B \textbf{98} 075422 (2018).
\bibitem{Iurovprb2023} A. Iurov, L. Zhemchuzhna, G. Gumbs, and D. Huang, Optical conductivity of gapped $\alpha-T_3$ materials with a deformed flat band, Phys. Rev. B \textbf{107} 195137 (2023).
\bibitem{Zhang2018} S.-H. Zhang, and W. Yang, Perfect transmission at oblique incidence by trigonal warping in graphene P-N junctions, Phys. Rev. B \textbf{97}, 035420 (2018).
\bibitem{Zhu2023} Y. Zhu, A. Merkel, L. Cao, Y. Zeng, S.Wan, T. Guo, Z. Su, S. Gao, H. Zeng, H. Zhang, and B. Assouar, Experimental
observation of super-Klein tunneling in phononic crystals, Appl. Phys. Lett. \textbf{122}, 211701 (2023).
\bibitem{Iurov2020}  A. Iurov, L. Zhemchuzhna, P. Fekete, G. Gumbs, and D. Huang, Klein tunneling of optically tunable Dirac particles with elliptical dispersions, Phys. Rev. Res. \textbf{2}, 043245 (2020).
\end{thebibliography}
\end{document}